\documentclass[5p,twocolumn,times]{elsarticle}

\usepackage{algorithm}
\usepackage{algorithmic}
\usepackage{amsmath}
\usepackage{amssymb}
\usepackage{color}
\usepackage[hang,flushmargin]{footmisc} 
\usepackage{graphicx}
\usepackage{listings}
\usepackage{subfig}

\definecolor{blue}{rgb}{0.098,0.357,0.675}
\definecolor{green}{rgb}{0.5,0.75,0.0}
\usepackage[
    pdftex,
    pdftitle={},
    pdfauthor={S.~K. Layton, A. Krishnan, L.~A. Barba},
    pdfpagemode={UseOutlines},
    bookmarks, bookmarksopen,bookmarksnumbered={True},
    colorlinks, linkcolor={blue},citecolor={blue},urlcolor={blue}
    ]{hyperref}

\lstdefinelanguage{CUDA}[]{C++}{
    morekeywords={__global__,__device__,__shared__,__syncthreads,threadIdx,blockIdx,float3,float4,rsqrtf},
}
\newcommand{\cpu}{\textsc{cpu}}
\newcommand{\gpu}{\textsc{gpu}}
\newcommand{\cuda}{\textsc{cuda}}
\newcommand{\cusp}{\textsl{Cusp}}
\newcommand{\NV}{\textsc{nvidia}}

\newcommand{\ibm}{\textsc{ibm}}
\newcommand{\cusparse}{\textsc{cusparse}}

\begin{document}

\begin{frontmatter}

\title{cuIBM -- A GPU-accelerated Immersed Boundary Method}

\author[bu]{S. K. Layton}
\ead{slayton@bu.edu}

\author[bu]{Anush Krishnan}
\ead{anush@bu.edu}

\author[bu]{L. A. Barba\corref{cor}}
\ead{labarba@bu.edu}

\cortext[cor]{Correspondence:  110 Cummington St, Boston MA 02215, (617) 353-3883}
\address[bu]{Department of Mechanical Engineering, Boston University, Boston, MA, 02215, USA.}

\begin{abstract}
A projection-based immersed boundary method is dominated by sparse linear algebra routines. Using the open-source {\cusp} library, we observe a speedup (with respect to a single {\cpu} core) which reflects the constraints of a bandwidth-dominated problem on the {\gpu}. Nevertheless, {\gpu}s offer the capacity to solve large problems on commodity hardware. This work includes validation and a convergence study of the {\gpu}-accelerated {\ibm}, and various optimizations.
\end{abstract}


\end{frontmatter}

\section{Introduction}

The immersed boundary method (\ibm) refers to a general class of techniques in computational fluid dynamics, characterized by the solution of the incompressible Navier-Stokes equations on a grid that does not conform to the immersed solid body present in the fluid. Conventional CFD techniques require the generation of a mesh that conforms to the geometry of the domain on which the equations are solved. In the immersed boundary method, the fluid is represented by an Eulerian grid (typically a Cartesian grid) and the solid boundary is represented by a collection of Lagrangian points. This has several advantages. Mesh generation is trivial, and simulations involving moving solid bodies and boundaries are made simpler. The Navier-Stokes equations are solved on the entire grid (including points within the solid), and the effect of the solid body is modelled by adding a singular force distribution $\mathbf{f}$ along the solid boundary which enforces the no-slip condition. The governing equations are,
\begin{subequations}
\begin{align}
\frac{\partial{\mathbf{u}}}{\partial{t}}+\mathbf{u}\cdot\boldsymbol{\nabla}\mathbf{u} & = -\boldsymbol{\nabla}{p}+\nu\Delta\mathbf{u} \nonumber \\
& \qquad +\int_s\mathbf{f}(\boldsymbol{\xi}(s,t))\delta(\boldsymbol{\xi}-\mathbf{x})ds \label{ns1} \\
\nabla\cdot\mathbf{u} & = 0 \label{ns2} \\
\mathbf{u}(\boldsymbol{\xi}(s,t)) & = \int_s\mathbf{u}(\mathbf{x})\delta(\mathbf{x}-\boldsymbol{\xi})d\mathbf{x} \nonumber \\
& = \mathbf{u}_B(\boldsymbol{\xi}(s,t)), \label{ns3}
\end{align}
\end{subequations}

\noindent where $\mathbf{u}_B$ is the velocity of the body at the boundary point locations. The different \ibm\ formulations use different techniques to calculate this forcing term.

The immersed boundary method was introduced in 1972 by C S Peskin \cite{Peskin1972} to model blood flow through the elastic membranes of the heart. The technique has since been extended to simulate rigid bodies, and the \ibm\ has seen renewed interest in recent times. Mittal and Iaccarino \cite{Mittal2005} provide a good overview, discussing various techniques that have been used to impose the no-slip condition on immersed boundaries.

Peskin \cite{Peskin1972} proposed modelling the body as a collection of springs, with boundary points placed at the equilibrium positions of the springs and allowed to move with the flow, and the force calculated using Hooke's law. The force is transferred from the solid to the fluid grid by using a well-chosen discretized approximation to the Dirac delta function. A second-order accurate extension to this method was developed in \cite{Lai2000}.

Goldstein \emph{et al.}\ \cite{Goldstein1993} generalised this technique by using a singular force at the boundary points, given by
${\mathbf{F}}(\mathbf{x}_s,t)=\alpha\int_{0}^{t}({\mathbf{u}}(\mathbf{x}_s,t)-{\mathbf{v}}(\mathbf{x}_s,t))dt+\beta({\mathbf{u}}(\mathbf{x}_s,t)-{\mathbf{v}}(\mathbf{x}_s,t))$,
where $\alpha$ and $\beta$ are constants that are chosen depending on the physical characteristics of the flow. This forcing function enforces the no-slip condition at the solid surface by means of a feedback loop. Saiki and Biringen \cite{Saiki1996} used this technique in conjunction with the finite difference method. Note that we obtain the same method as \cite{Lai2000} by setting $\beta$ to zero.

These methods are easy to implement but have some shortcomings. The no-slip condition is not directly applied, and it takes some time for the system to attain it. The algorithms also depend on \emph{ad hoc} parameters that need to be chosen carefully depending on the physics of the flow and the desired accuracy. The typically large values of the parameters also place a very strict restriction on the time step due to stability constraints.

Mohd-Yusof \cite{Mohd-Yusof1998} formulated a method in which the forcing term is directly obtained from the discretised Navier-Stokes equations by assuming no-slip at the solid boundaries. The velocity at the grid points near the boundary is estimated via interpolation and the forcing term is calculated by
\begin{equation}
\mathbf{f} = \rho\left(\frac{\partial{\mathbf{u}}}{\partial{t}}+\mathbf{u}\cdot{\nabla}\mathbf{u}\right)+\nabla{p}-\mu\Delta\mathbf{u}.
\end{equation}
This technique was implemented by Fadlun \emph{et al.}\ \cite{Fadlun2000}. An implicit scheme to calculate the diffusion and the forcing term allowed large CFL numbers of the order of $10^{-1}$ to be used. This method does not exactly satisfy the no-slip condition, but the errors were shown to be small.  Kim \emph{et al.}\ \cite{Kim2001} used a similar idea, with the addition of a mass source at the boundaries. The forcing term and mass sources were calculated explicitly, which placed a stricter restriction on the time step.

Uhlmann \cite{Uhlmann2005} reported that the method in \cite{Fadlun2000} produced a boundary force that was not smooth in time when it was used to solve flows with moving boundaries. To fix this, he recommended calculating the singular force distribution at the Lagrangian body points and then distributing the force on the Eulerian grid by using a discrete delta function, rather than calculating the forcing term directly. This method too does not satisfy the no-slip condition exactly and has the stricter time step restriction, but eliminated the force oscillations.

The \emph{immersed boundary projection method} was introduced by Taira and Colonius \cite{Taira2007}, based on an extension of the projection method \cite{Chorin1968, Kim1985, Perot1993} used to solve the incompressible Navier-Stokes equations. Using the idea that the pressure in incompressible flows acts as a Lagrangian multiplier to ensure the zero-divergence condition, they  similarly consider the forcing term as Lagrangian multiplier that ensures the no-slip condition on the immersed boundary. During the projection step, the velocity field is adjusted such that both these constraints are satisfied exactly. An implicit scheme used for diffusion allows the usage of relatively large time steps, with CFL numbers as high as $1$. They later used a nullspace approach that sped up the simulation by up to two orders of magnitude \cite{Colonius2008}. This approach essentially uses a spectral method to solve the equations cast in the velocity-vorticity formulation.

The techniques described so far make use of some kind of interpolation to transfer the singular forces at the boundary to the Eulerian grid on which the Navier-Stokes equations are solved. This causes the boundary to appear diffused and affects the accuracy of the boundary layer. Boundary layers need to be resolved more accurately at higher Reynolds numbers, and so a number of \emph{sharp-interface} methods have been developed to handle this problem.
Among them is the cut-cell method. Cells near the boundary are cut to shapes that conform to the body, and the rest of the grid is Cartesian. The Navier-Stokes equations are solved using the finite volume method in all the cells that contain the fluid and no forcing term is used. Both local and global conservation of properties in the domain can be ensured. This technique has been used by Ye \emph{et al.}\ \cite{Ye1999}, Udaykumar \emph{et al.}\ \cite{Udaykumar1997, Udaykumar1999, Udaykumar2001}, Mittal \emph{et al.}\ \cite{Mittal2003, Mittal2002} and Marella \emph{et al.}\ \cite{Marella2005}. Extending this procedure to three dimensions is non-trivial due to the complex shapes that can be formed by cutting.

Another sharp-interface method is known as the ghost-cell method. Here, an interpolation scheme is used at the interface to calculate flow properties at the grid points, implicitly taking into account the boundary conditions. This technique has been used by Majumdar \emph{et al.}\ \cite{Majumdar2001}, Iaccarino and Verzicco \cite{Iaccarino2003}, Ghias \emph{et al.}\ \cite{Ghias2004} and Kalitzin \emph{et al.}\ \cite{Kalitzin2003}, the last two of whom carried out turbulence simulations.

In the present work, we use the algorithm presented in \cite{Taira2007} for the solution of two-dimensional incompressible viscous flows with immersed boundaries, which is explained in detail in \S\ref{sec:ibpm}. We have produced an implementation on \gpu\ hardware, studying techniques that provide good performance in this multi-threaded architecture, which are described in some detail. The various challenges that an efficient implementation on \gpu\ pose are representative of those that practitioners of other CFD methods would face. Thus, we hope to contribute to the ongoing investigations on the use of \gpu\ hardware for computational fluid dynamics. To our knowledge, the {\ibm} has not previously been implemented on the \gpu. The perspective of doing so is the capacity of solving large three-dimensional moving boundary problems on commodity hardware.

\section{Immersed Boundary Projection Method}\label{sec:ibpm}

We implement the method proposed by Taira \& Colonius \cite{Taira2007}. The Navier-Stokes equations \eqref{ns1}--\eqref{ns3} are discretised on a staggered cartesian grid. The grid can be non-uniform, and it is advantageous to stretch the grid away from the high vorticity regions to reduce the number of grid points without sacrificing accuracy. We obtain the following set of algebraic equations:

\begin{align}
	\hat{A}u^{n+1} - \hat{r}^n & = - \hat{G}\phi + \hat{bc}_1 + \hat{H}f \notag \\
	\hat{D}u^{n+1} & = bc_2
	\label{disns} \\
	\hat{E}u^{n+1} & = u_B^{n+1}, \notag
\end{align}

which can be written as:

\begin{equation}\label{eqn:disnsmat}
	\left(
		\begin{array}{ccc}
			\hat{A} & \hat{G} & \hat{H} \\
			\hat{D} & 0 & 0 \\
			\hat{E} & 0 & 0\end{array}
	\right)
	\left(
		\begin{array}{c}
			u^{n+1} \\
			\phi \\
			f
		\end{array}
	\right)
	=
	\left(
		\begin{array}{c}
			\hat{r}^n \\
			0 \\
			u_B^{n+1}
		\end{array}
	\right)
	+
	\left(
		\begin{array}{c}
			\hat{bc}_1 \\
			-bc_2 \\
			0
		\end{array}
	\right)
\end{equation}

\noindent
where $\phi$ and $f$ are vectors whose contents are the pressure and the components of the forces at the immersed boundary points respectively. The velocity at the current time step $u^n$ is known. $\hat{bc}_1$ and $bc_2$ are obtained from the boundary conditions on the velocity.

$\hat{H}$ and $\hat{E}$ are the regularisation and interpolation matrices respectively, which are used to transfer values of the flow variables between the Eulerian and Lagrangian grids. The interpolation in \eqref{ns3} can be discretised as:

\begin{equation}
	u_k = \sum_{i}{u_i}d(x_i-\xi_k)d(y_i-\eta_k)\Delta{x}\Delta{y},
\end{equation}

\noindent
where $u_k$ is the velocity at a point $(\xi_k,\eta_k)$ on the immersed boundary. $u_k$ is calculated by convolving the velocities on the Eulerian grid $u_i$ at locations $(x_i,y_i)$ with a discrete two-dimensional delta function. The discrete delta function is a product of smoothed one-dimensional delta functions $d_h(r)$ along each Cartesian direction. We choose the one used by Roma et al  \cite{Roma1999}:

\begin{equation}
	d_{h}(r) = \left \{
\begin{array}{l}
	\frac{1}{6h}\left(5-3\frac{|r|}{h}-\sqrt{-3\left(1-\frac{|r|}{h}\right)^2+1}\right), \quad 0.5<\frac{|r|}{h}{\le}1.5 \notag \\
	\frac{1}{3h}\left(1+\sqrt{-3\left(\frac{|r|}{h}\right)^2+1}\right), \quad 0<\frac{|r|}{h}{\le}0.5 \notag \\
	0, \quad otherwise,
\end{array}
\right .
\end{equation}

\noindent
where $h$ is the cell width. Use of the discrete delta function requires the grid to be uniform near the immersed boundary. Sufficient number of boundary points need to be chosen to prevent flow leakage, and the spacing between boundary points should be around the same as the cell width of the Eulerian grid.

From the above, we can calculate the elements of matrix $\hat{E}$:

\begin{equation}
	\hat{E}_{k,i}=\Delta{x}\Delta{y}d(x_i-\xi_k)d(y_i-\eta_k).
\end{equation}

The matrix $\hat{H}$ is similarly determined by discretising the forcing term in \eqref{ns1}.

The explicit second-order Adams-Bashforth scheme is used to discretise the convection terms and the Crank-Nicolson scheme is used for diffusion. All spatial derivatives are calculated using central differences. Note that no explicit pressure boundary conditions need to be specified. The pressure and body forces are  calculated implicitly. The above system of equations can be solved to obtain the velocity field at time step $n+1$, the pressure (to a constant) and the body forces. But the left-hand side matrix is indefinite, and solving the system directly is ill-advised.

By performing appropriate transformations (see Appendix of \cite{Taira2007} for details), we can show that the above system is equivalent to:
\begin{equation}\label{eqn:system_full}
	\left(
		\begin{array}{ccc}
			A & G & E^T \\
			G^T & 0 & 0 \\
			E & 0 & 0\end{array}
	\right)
	\left(
		\begin{array}{c}
			q^{n+1} \\
			\phi \\
			\tilde{f}
		\end{array}
	\right)
	=
	\left(
		\begin{array}{c}
			r^n \\
			0 \\
			u_B^{n+1}
		\end{array}
	\right)
	+
	\left(
		\begin{array}{c}
			bc_1 \\
			-bc_2 \\
			0
		\end{array}
	\right)
\end{equation}

where $q^{n+1}$ is the momentum flux at each cell boundary, and the sub-matrices are transformed versions of those in \eqref{eqn:disnsmat}. We can rewrite the above by combining some of the sub-matrices in the following manner:

\begin{equation}
	Q =
	\left[
		\begin{array}{cc}
			G & E^T
		\end{array}
	\right],~~
	\lambda =
	\left(
		\begin{array}{c}
			\phi \\
			\tilde{f}
		\end{array}
	\right),~~
	r_1=r^n+bc_1,~~
	r_2=
	\left(
		\begin{array}{c}
			-bc_2 \\
			u_B^{n+1}
		\end{array}
	\right)
\end{equation}

\noindent
which gives us

\begin{equation}\label{newsystem}
	\left(
		\begin{array}{ccc}
			A & Q \\
			Q^T & 0
		\end{array}
	\right)
	\left(
		\begin{array}{c}
			q^{n+1} \\
			\lambda
		\end{array}
	\right)
	=
	\left(
		\begin{array}{c}
			r_1 \\
			r_2
		\end{array}
	\right)
\end{equation}

Applying the approximate factorisation described in \cite{Perot1993} to the above system, we obtain the following set of equations which can be solved to give us the velocity distribution at time step $n+1$:

%
%
\begin{subequations}
\begin{align}
	A q^{*} & =r_1 \label{ibpm1}\\
	Q^T B^N Q \lambda & = Q^T q^{*} - r_2 \label{ibpm2}\\
	q^{n+1} & = q^* - B^N Q \lambda \label{ibpm3}
\end{align}
\end{subequations}

where $B^N$ is an $N^{th}$ order approximation of $A^{-1}$. This must be taken into account while estimating the overall time accuracy of the method.

This factorisation is very advantageous as the two linear systems \eqref{ibpm1} and \eqref{ibpm2} that we now need to solve can be made positive definite, and can be solved efficiently using the conjugate gradient method. In the absence of an immersed boundary, this set of equations is the same as that solved in the traditional fractional step method or projection method \cite{Perot1993, Kim1985, Chorin1968}. The final equation \eqref{ibpm3} simultaneously ensures a divergence-free velocity field and the no-slip condition in the next time step. By virtue of the above formulation, no special boundary conditions need to be derived for $q^*$ or $\lambda$. As is usual, one value of pressure needs to be pinned to obtain a unique solution since the left hand side matrix $Q^T B^N Q$ in \eqref{ibpm3} has one eigenvalue which is zero.

\section{Assessment of iterative solvers for the {\ibm}}

Like a majority of numerical methods used in computational mechanics, the \ibm\ requires tools for sparse linear algebra. The matrices $A$, $Q$ and $Q^T$ are sparse, the vectors $q^{n+1}$ and $\lambda$ are dense, and we need to operate on these objects via matrix-vector, matrix-matrix, and even a triple-matrix multiplication. To take advantage of the {\gpu}, we need efficient means of both representing and operating on these matrices and vectors on the device. Currently, there are two tools available for this purpose: {\cusparse}, part of {\NV}'s {\cuda}, or the external library, {\cusp}. The {\cusp} library is being developed by several {\NV} employees with minimal software dependencies and released freely under an open-source license. We chose to use the {\cusp} library for several reasons: it is actively developed and separate from the main {\cuda} distribution, allowing for faster addition of new features (such as new pre-conditioners, solvers, \emph{etc.}); and, all objects/methods from the library are usable on both {\cpu} and {\gpu}. This allows us the flexibility to, for example, perform branching-heavy code on the {\cpu}, before trivially transferring to the device and running (for instance) a linear solve, where it will be significantly faster. It also allows us to maintain both a {\cpu} and {\gpu} code with less effort.

\begin{figure*}
\centering
	\subfloat[Timing breakdown]{
		\includegraphics[width=0.45\textwidth]{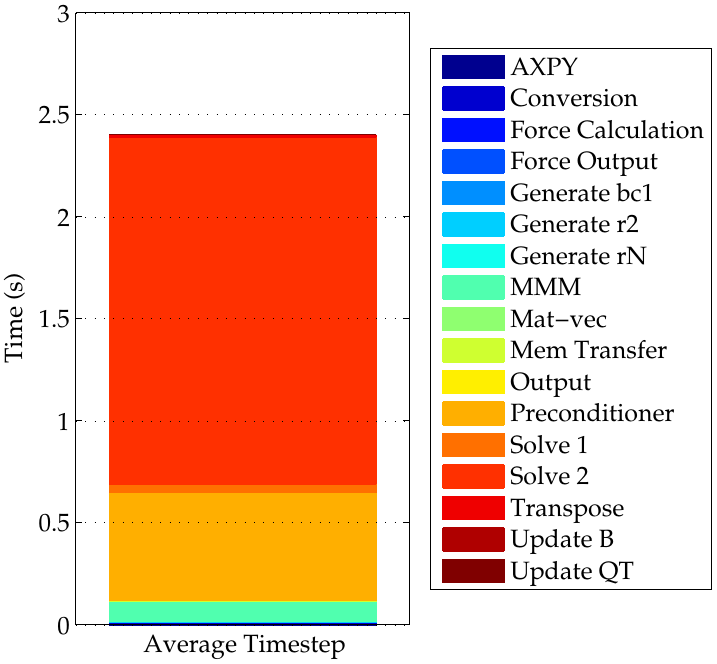}\label{fig:breakdown}}
	\subfloat[Solving linear equations]{
		\includegraphics[width=0.45\textwidth]{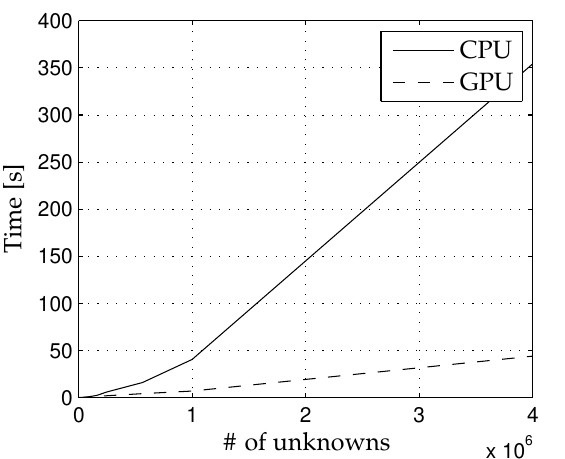}\label{fig:solve_speed}}
		\caption{\small (a)  Timing breakdown for a flapping airfoil at $Re=75$ using the \gpu\ code. (b) Comparison of time taken to solve a system of linear equations $Ax=b$ on the {\cpu} and {\gpu}. $A$ is chosen as the standard $5$-pt Poisson stencil.}
\end{figure*}

To illustrate the importance of having efficient linear algebra methods, we show in Figure \ref{fig:breakdown} a breakdown of the timings from an example run on the {\gpu} the with the \ibm\ ($4000$ time steps of a flapping airfoil at $Re = 75$). The mesh consists of $930\times 654$ cells, resulting in systems of over $600,000$ unknowns. Even in this moderately sized test, the runtime is dominated by the solution of the coupled linear system for pressure and forcing terms, denoted by `Solve 2'.
Speeding up this linear solve is the major motivation for using the {\gpu}.

At this point, it is interesting to consider the potential benefits of using the {\gpu} for our linear systems. Sparse linear systems are a well known example of a bandwidth-limited problem\,---\,the potential speedup from using a {\gpu} will at best be the ratio of the bandwidths of the {\cpu} and {\gpu} used. For the Intel Xeon X5650 processors in our workstation, Intel quotes a maximum bandwidth of $32$ GB/s\footnote{\href{http://ark.intel.com/products/47922/Intel-Xeon-Processor-X5650-(12M-Cache-2_66-GHz-6_40-GTs-Intel-QPI)}{http://ark.intel.com/products/47922/Intel-Xeon-Processor-X5650-(12M-Cache-2\_66-GHz-6\_40-GTs-Intel-QPI)}}, while {\NV} quotes $144$ GB/s for the Tesla C2050 cards we use\footnote{\href{http://www.nvidia.com/docs/IO/43395/NV_DS_Tesla_C2050_C2070_jul10_lores.pdf}{http://www.nvidia.com/docs/IO/43395/NV\_DS\_Tesla\_C2050\_C2070\_jul10\_lores.pdf}}. Using these numbers, we can see that the best-case speedup should be $~4.5\times$ for a purely bandwidth-bound problem, while for computationally bound problems the speedup can be much higher. In practice, even the worst case speedup is definitely worthwhile, certainly enough to justify the extra complexity involved in using the {\gpu}.

Figure \ref{fig:solve_speed} shows a timing comparison between the {\cpu} and {\gpu} using {\cusp}'s conjugate gradient solver. The system being solved in this case is given by a traditional 5-point Poisson stencil, which while not directly used in the \ibm\ code, gives a good measure of relative performance that can potentially be obtained. The plot shows the wall-clock time required to solve to a relative accuracy of $10^{-5}$ for numbers of unknowns ranging from $2500$ to $4\times 10^6$. For large systems, the {\gpu} solve is significantly faster, with a speedup of $8\times$ for the largest system shown. This indicates that even though our actual system may be much harder to solve, the potential speedup is significant.

\begin{figure}[h]
\centering
	\includegraphics[width=0.45\textwidth]{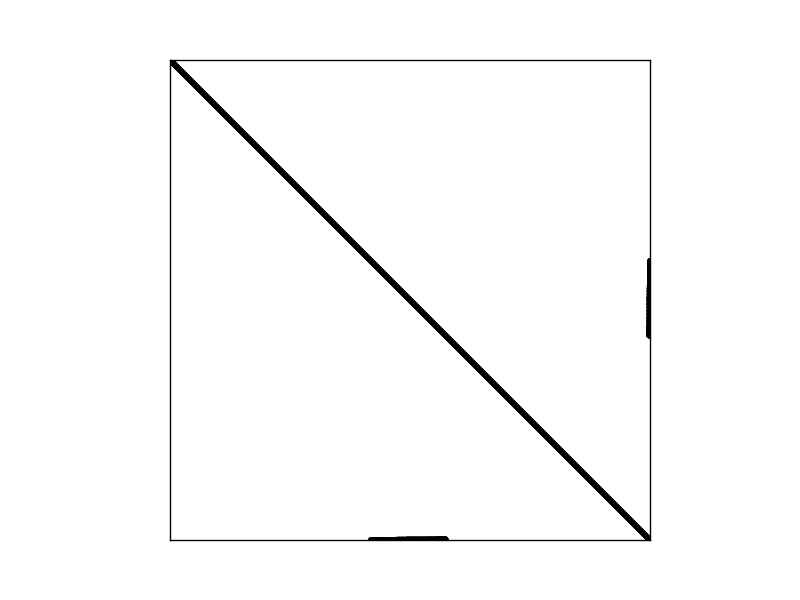}
	\caption{\small Example of sparsity pattern for $Q^{T}B^{N}Q$ matrix, showing the coupling terms in the right and lower sections of the matrix}
\label{fig:sparsity}
\end{figure}

For the particular linear system we're interested in, we also have to take into account that it has an unusual non-zero structure, shown in Figure \ref{fig:sparsity}, due to the coupled nature of the variables being solved for (pressure and forcing terms). This coupled nature leads to conflicting approaches being optimal for different sections of the matrix. The pressure terms are a standard 5-point Poisson stencil (to be extended to 7-point in 3D), and as such would be a good candidate for a hierarchical method using a smoother, such as algebraic multigrid (AMG). However, the forcing terms and coupling sections of the matrix would be unsuitable for this kind of approach. Instead, a standard solver such as CG (Conjugate Gradient) would be better suited.

To tackle this difficulty, we can choose one of two possible approaches: \emph{(i)} use a suitable pre-conditioner for the system to make it better suited for solution with a standard iterative solver, most likely CG, or \emph{(ii)} investigate other forms of solvers, such as AMG.
While option \emph{(ii)} could potentially have the greatest effect on both convergence rate and total time to solution, our choices are somewhat limited, given the tools we are using. Either we rely on the smoothed aggregation AMG that is part of {\cusp}, or we would need to develop our own from scratch---at this time there are no other \gpu-enabled AMG codes available for public use. 

To provide further insight into our options, we can test representative matrices from different applications of our code (with differing complexities of immersed boundaries) and compare a standard CG solver against the form of AMG available in {\cusp}, an aggregative AMG. The CG solver will be used both without a preconditioner, as well as using the diagonal and smoothed-aggregation preconditioners from {\cusp}. For each problem, we show the wall-clock time taken to solve the system to a relative accuracy of $10^{-5}$, the same accuracy used in previous tests. A summary of the matrices produced from each system can be seen in Table \ref{tab:test_details}, and the timing results are shown in Figure \ref{fig:solver_speeds}.

From these results, it can be plainly seen that PCG with smoothed-aggregation AMG as the preconditioner is the fastest on all tests, and so this is the combination of solver and preconditioner used in all our subsequent runs.

\begin{figure*}
\centering
	\includegraphics[width=0.6\textwidth]{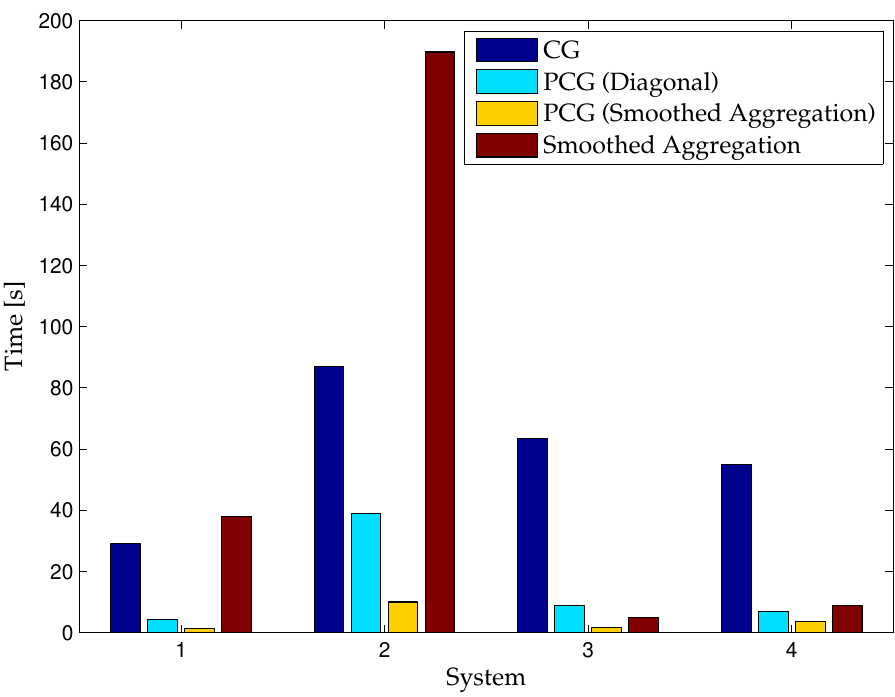}
	\caption{\small Comparison of {\gpu} solvers on real test problems - Conjugate Gradient, Preconditioned Conjugate Gradient with diagonal and Smoothed Aggregation preconditioners, and a full solve using Smoothed Aggregation}
	\label{fig:solver_speeds}
\label{fig:solver_speeds}
\end{figure*}

\begin{table*}
\begin{center}
\small
	\begin{tabular}{| c | c | c | c | c | }
	\hline
		System \# & Domain Size & \# Bdy Points & \# Unknowns & \# Non-zeros \\
	\hline
		1 & $330\times330$ & 158 & 109216 & 554752 \\
		2 & $986\times986$ & 786 & 973768 & 4914520 \\
		3 & $930\times654$ & 101 & 608422 & 3045406 \\
		4 & $690\times690$ & 474 & 477048 & 2412448 \\
	\hline
	\end{tabular}
	\caption{\small Systems $1$ and $2$ correspond to flow over a cylinder at $Re = 40$ and $3000$ respectively, system $3$ is for a flapping-wing calculation, and system $4$ is from a synthetic test with $3$ moving cylinders.}
	\label{tab:test_details}
\end{center}
\end{table*}

Our choice of tools allows us to easily perform all sparse linear algebra operations on the {\gpu}. On the other hand, there are parts of the algorithm that cannot easily be expressed using linear algebra, such as generating the convection term using a finite-difference stencil and applying boundary conditions to the velocities (which involves modifying select values of appropriate arrays). One possible way of performing these actions is to transfer data from the {\gpu}, perform the calculations on the {\cpu} and transfer the modified vector(s) back to the {\gpu} every time step\,---\,but this incurs a prohibitively high cost in memory transfers. The alternative, which we have done, is to use custom-written {\cuda} kernels utilizing all appropriate techniques, including the use of shared memory, to perform these operations on the {\gpu}. This requires access to the underlying data from the {\cusp} data structures, which can be done easily using the \textsl{Thrust} library, on which {\cusp} was built. 

\section{Strategy for \gpu\ implementation}

Here we discuss some important implementation details that are used to decrease the total runtime, or to decrease the amount of memory needed, or both.

\begin{itemize}

\item \emph{Keep everything on the {\gpu}: } To reduce costly memory transfers, everything that can be performed on the {\gpu} is, regardless of its suitability. For instance, the code to enforce boundary conditions operates on very few values, has almost no computational complexity and is highly divergent based on the boundary conditions chosen (\emph{i.e.}, Dirichlet / Neumann, \emph{etc.}). Thus, it will always perform rather badly on the device, giving essentially no speedup. However, it would take significantly longer to transfer the necessary data back to the host and perform the operation there. This means taking the time to write and test this kernel gives us an overall speedup, due to this elimination of transfers. This is only one example, but all but one element of the total algorithm has been transferred, yielding significant savings in the total run time. The final element still on the {\cpu} involves updating the forcing terms in $Q$, and as such involves a very small number of entries compared to any other array. Thus, this transfer ends up being an inconsequential part of the total algorithm.

\item \emph{Triple Matrix Product: } To generate the left-hand side of equation (\ref{ibpm2}), we need to multiply 3 sparse matrices, $Q^{T}$, $B^{N}$ and $Q$. To do this within the current {\cusp} framework, we would need to perform two separate matrix-matrix multiplies, the first for $Q^{T}B^{N}$ and then the result of this multiplied by $Q$. This method involves creating a temporary matrix in memory, and for higher-order approximations of $B^{N}$, say, $3^{rd}$-order, this temporary can be quite large, in the order of $700\mathrm{MB}$. {\cusp} performs matrix-matrix multiplies in the following form (using Matlab notation, where $A\verb|[range,range]|$ denotes a sub matrix of $A$, and the `$\verb|:|$' symbol denotes all columns/rows depending on context; thus, $A\verb|[slice,:]|$ refers to all columns in a range of rows denoted by $\verb|slice|$)

\begin{algorithmic}
	\REQUIRE Input sparse matrices $A$, $B$, result matrix $C$
	\STATE Split $A$ into slices $\verb|A[slice,:]|$ s.t. each $\verb|A[slice,:]|\cdot \verb|B|$ fits in device memory
	\FOR{$\forall$ slices}
		\STATE $\verb|slice| \gets \verb|A[slice,:]|\cdot \verb|B|$
		\STATE add slice to list of slices
	\ENDFOR
	\STATE Assemble $C$ from all computed slices
\end{algorithmic}

\noindent
The multiplication of $\verb|A[slice,:]|\cdot \verb|B|$ is here being performed by a helper function within \cusp\ (\verb|cusp::detail::device::spmm_coo_helper|).

We propose a routine to perform a triple matrix product of the form $D \gets A\cdot B\cdot C$ by using this helper function repeatedly to ensure the full intermediate product need not be calculated. We do this by realizing the following:
\begin{eqnarray*}
	\verb|temp_slice| & = & \verb|A[slice,:]|\cdot \verb|B| \\
	\verb|D[slice,:]| & = & \verb|temp_slice|\cdot \verb|C|.
\end{eqnarray*}

Thus, we can form slices of the final result while only storing a slice as an intermediate by applying the helper function twice in succession. This alleviates the need to create and store the full intermediate matrix. While at first glance the intermediate, $B^{N}Q$ might be computed once and used in both (\ref{ibpm2}) and (\ref{ibpm3}), we explain below how this term is unnecessary in (\ref{ibpm3}).

\item \emph{Use optimized routines where available: } In equation (\ref{ibpm3}), we have to calculate the product $B^{N}Q\lambda$, where $B^{N}$ and $Q$ are sparse matrices, and $\lambda$ is a dense vector. If we implement this naively, we perform the matrix-matrix product of $B^{N}Q$ then multiply this with the vector $\lambda$. However, the Sparse matrix-vector product (SpMV) in {\cusp} has been optimized to a significant state, as demonstrated by the co-authors of {\cusp}\cite{BellGarland2009}. Therefore, we prefer to implement this operation as a pair of SpMV operations, first $Q\lambda$ resulting in a vector, then the result of this with $B^{N}$, exchanging the matrix-matrix product with a SpMV. This also reduces the amount of memory needed, as we have no need to store the extra matrix obtained by $B^{N}Q$.

\item \emph{Re-use the hierarchy generated by the smoothed-aggregation preconditioner: } The most expensive part of our second solve, Equation \eqref{ibpm2} for moving bodies, is the generation of the hierarchy for the smoothed aggregation preconditioner. When bodies are stationary, we have no need to recompute this at each time step, and can use the same hierarchy for the entire simulation, effectively amortising the high cost over the total run. When we have moving bodies, this hierarchy should be recalculated every time step. We propose to instead re-use the hierarchy for several time steps. To do so, we need to investigate the balance of not performing the expensive calculation every time step, but potentially having a lower convergence rate for the steps where the hierarchy is re-used. For a flapping airfoil, Figure \ref{fig:npc_steps} shows the effect on the total run time of recalculating the preconditioner every $n$ time steps, clearly showing that in order to reduce run time for this problem, we should recalculate the preconditioner every 2 or 3 time steps.

\begin{figure}[h]
\centering
	\includegraphics[width=0.45\textwidth]{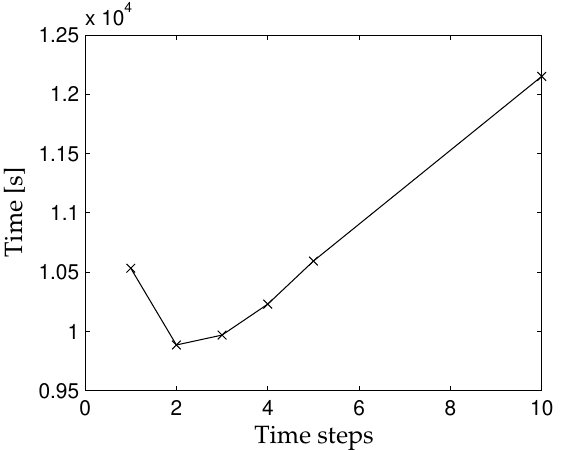}
	\caption{\small Timings for a flapping airfoil run, re-calculating the Smoothed Aggregation precondtioner every number of time steps.}
	\label{fig:npc_steps}
\end{figure}

\end{itemize}

\section{Validation}

\subsection{Couette flow between concentric cylinders}

As a validation test, we calculate the flow between two concentric cylinders of radius $r_i=0.5$ and $r_o=1$ centered at the origin. The outer cylinder is held stationary while the inner cylinder is impulsively rotated from rest with an angular velocity of $\Omega=0.5$. The cylinders are contained in a square stationary box of side $1.5$ centered at the origin. The fluid in the entire domain is initially at rest and the calculations were carried out for kinematic viscosity $\nu=0.03$.

The steady-state analytical solution for this flow is known. The velocity distribution in the interior of the inner cylinder is the same as for solid body rotation and the azimuthal velocity between the two cylinders is given by:
\begin{equation}
u_{\theta}(r)=\Omega{r_i}\frac{\left({r_o}/{r}-{r}/{r_o}\right)}{\left({r_o}/{r_i}-{r_i}/{r_o}\right)}.
\end{equation}

\begin{figure}[h]
\centering
	\subfloat[]{
		\includegraphics[width=0.45\textwidth]{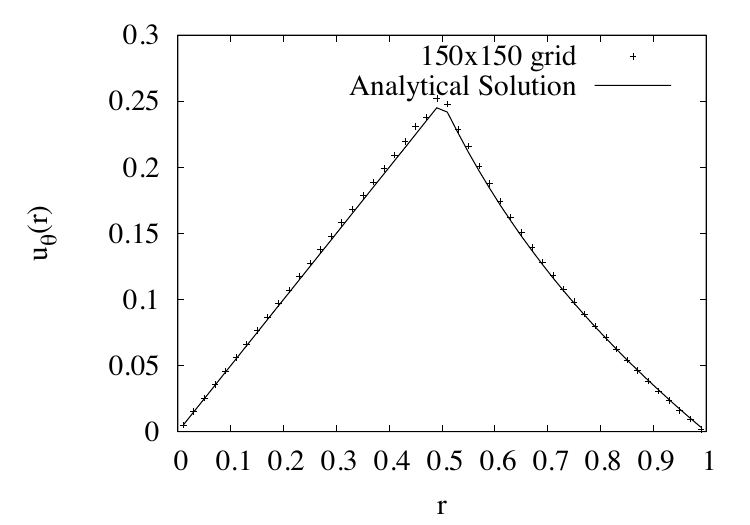}\label{fig:validate}}\\
	\subfloat[]{
		\includegraphics[width=0.45\textwidth]{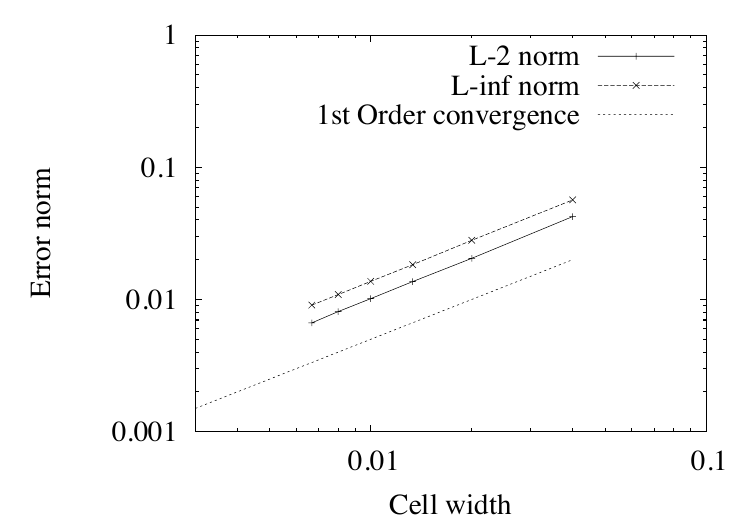}\label{fig:spatial}}
	\caption{\small (a) Comparison of the numerical solution on a $150\times 150$ grid with the analytical solution and (b) convergence study, showing errors for different grid sizes.}
\end{figure}

We compared this to the numerical solution for six different grid sizes ranging from $75\times{75}$ to $450\times{450}$. Table \ref{table:temporal} shows the $L^2$ and $L^\infty$ norms of the relative errors and Figure \ref{fig:spatial} shows that the scheme is first-order accurate in space, as expected for the \ibm\ formulation we used.

\begin{table}
\begin{center}
\small
    \begin{tabular}{ | c | c | c | }
 \hline
     & Order of convergence & Order of convergence \\
    Time & ($N=1$) & ($N=3$) \\ \hline
    0.8 & 0.97 & 2.67 \\ \hline
    2 & 0.99 & 2.85 \\ \hline
    4 & 0.93 & 2.73 \\ \hline
    8 & 0.97 & 2.83 \\ \hline
    \end{tabular}
\end{center}
    \vspace{-4mm}\caption{\small Calculated order of convergence at different times for Couette-flow validation.}
    \label{table:temporal}
\end{table}

To verify the temporal order of convergence, we ran a simulation from $t=0$ to $t=8$ on a $151\times 151$ grid, using different time steps ($\Delta{t}=0.01,~0.005~\text{and}~0.0025$). Both first- and third-order accurate expansions of $B^N$ were used and the calculated orders of convergence (using the $L^2$ norms of the differences in the solutions) at various times have been summarised in Table \ref{table:temporal}, and are as expected.

\subsection{Flow over an impulsively started cylinder}

We also carried out computations to simulate flow over an impulsively started circular cylinder at Reynolds numbers $40$, $550$ and $3000$.  The cylinder is of diameter $d=1$ centered at the origin and is placed in an external flow with freestream velocity $u_{\infty}=1$. The domain considered was square with each side of length 30, centered at the origin. The fluid flows from left to right, and the velocity on the left, top and bottom edges was set to be the freestream velocity. A convective boundary condition ($\frac{\partial{u}}{\partial{t}}+u_{\infty}\frac{\partial{u}}{\partial{x}}=0$) was used on the right edge. The initial velocity is uniform throughout the domain. The minimum cell widths used near the solid boundaries were 0.02, 0.01 and 0.004 respectively for the three cases. Away from the body, the grid is an exponential stretched grid. More information about the grids is given in Table \ref{table:gridinfo}.

\begin{figure*}
\begin{center}
	\subfloat[]
		{\includegraphics[width=0.37\textwidth]{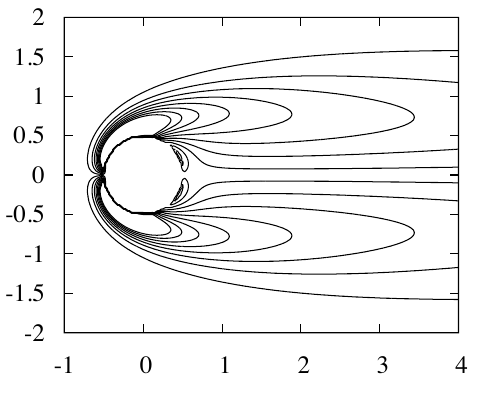}}
	\subfloat[]
		{\includegraphics[width=0.4\textwidth]{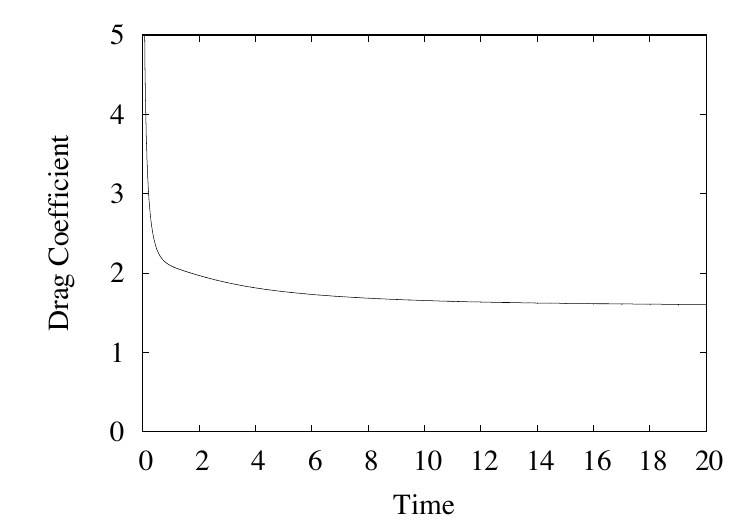}}
		\vspace{-3mm}\caption {\small (a) Steady state vorticity field and (b) time varying drag coefficient (b) for external flow over a circular cylinder at Reynolds number 40. The contour lines in (a) are drawn from -3 to 3 in steps of 0.4.}
		\label{fig:Re40}
	\subfloat[]
		{\includegraphics[width=0.37\textwidth]{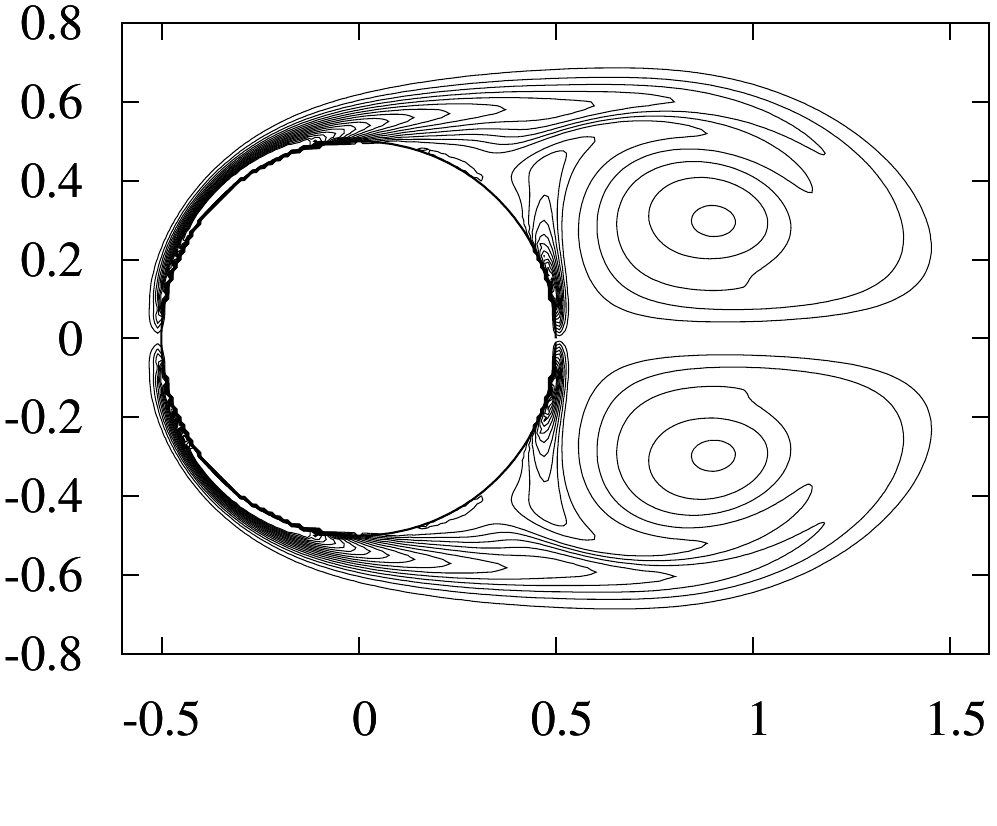}}
	\qquad
	\subfloat[]
		{\includegraphics[width=0.4\textwidth]{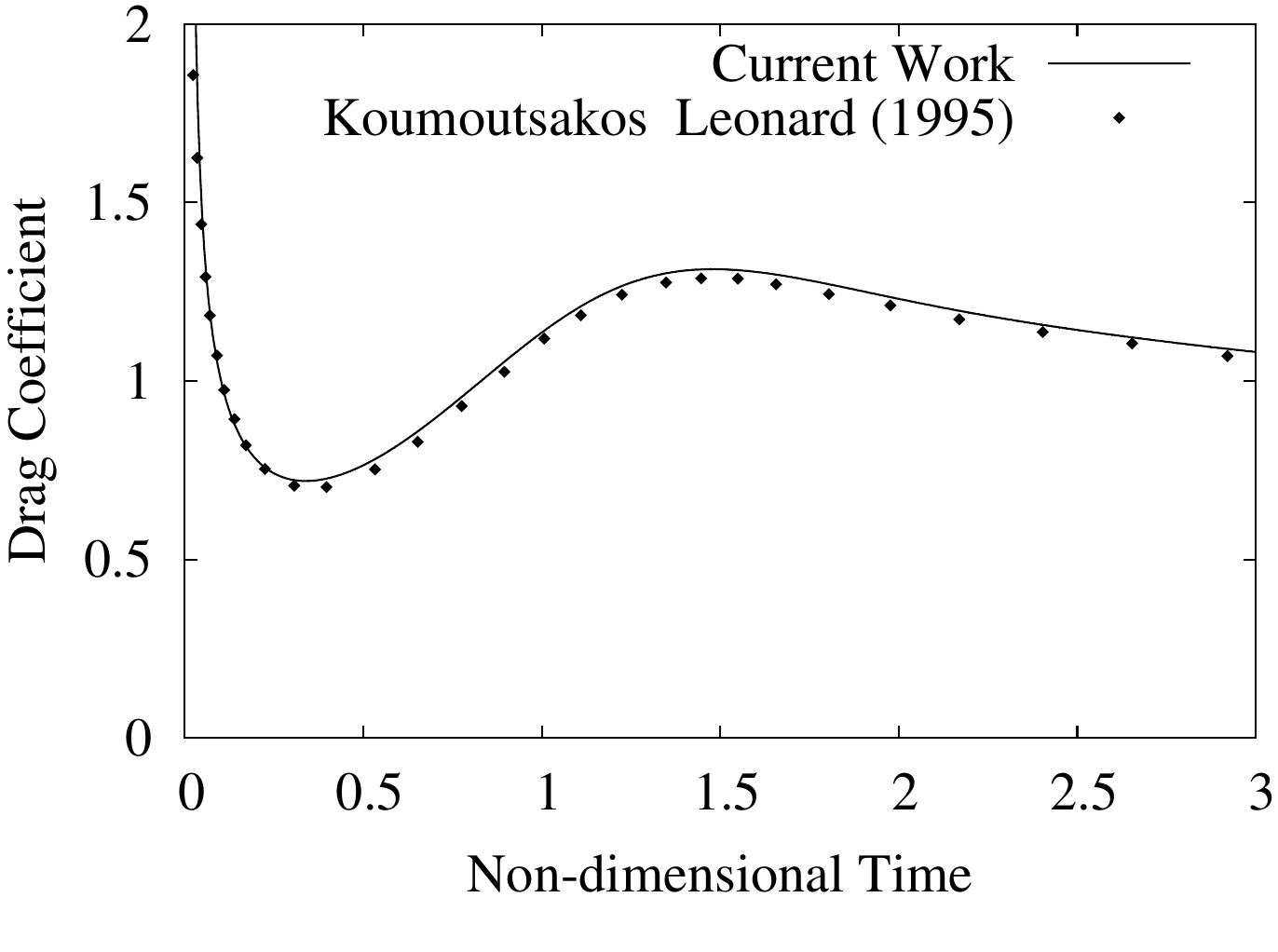}}
	\vspace{-3mm}\caption {\small (a) Vorticity field after non-dimensional time 3.0 and (b) time varying drag coefficient for external flow over a circular cylinder at Reynolds number 550. The contour lines in (a) are drawn from -32 to 32 in steps of 2, excluding the zero contour.}
	\label{fig:Re550}
	
	\subfloat[]
		{\includegraphics[width=0.37\textwidth]{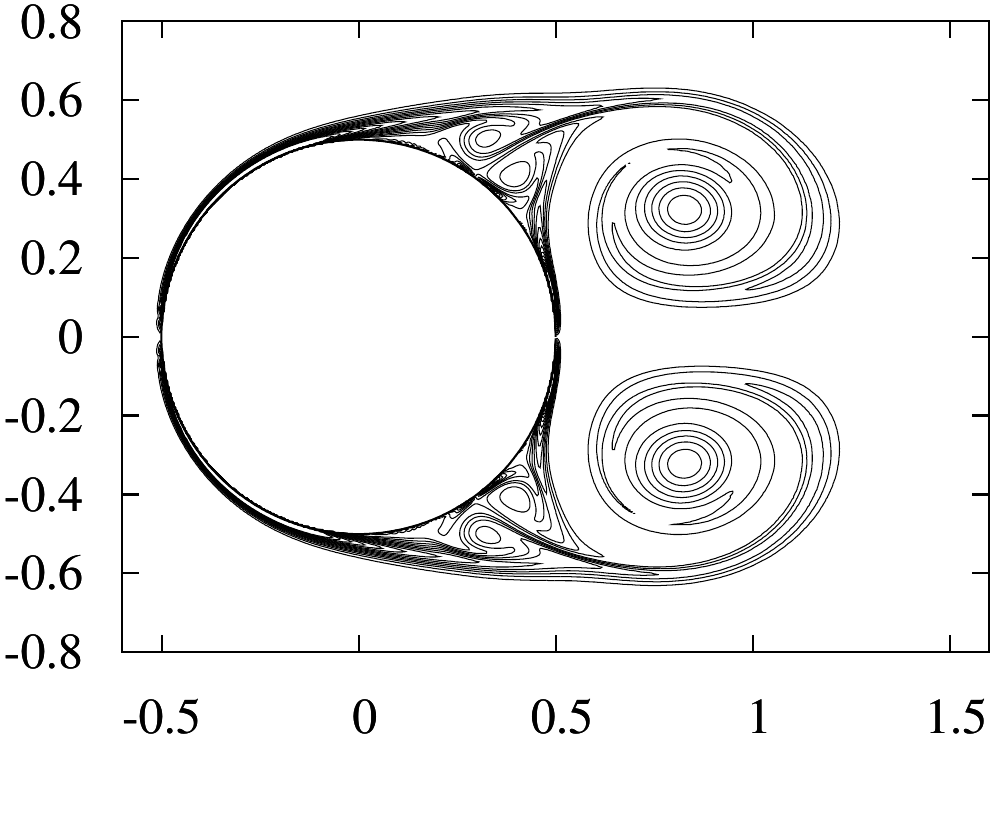}}
	\qquad
	\subfloat[]
		{\includegraphics[width=0.4\textwidth]{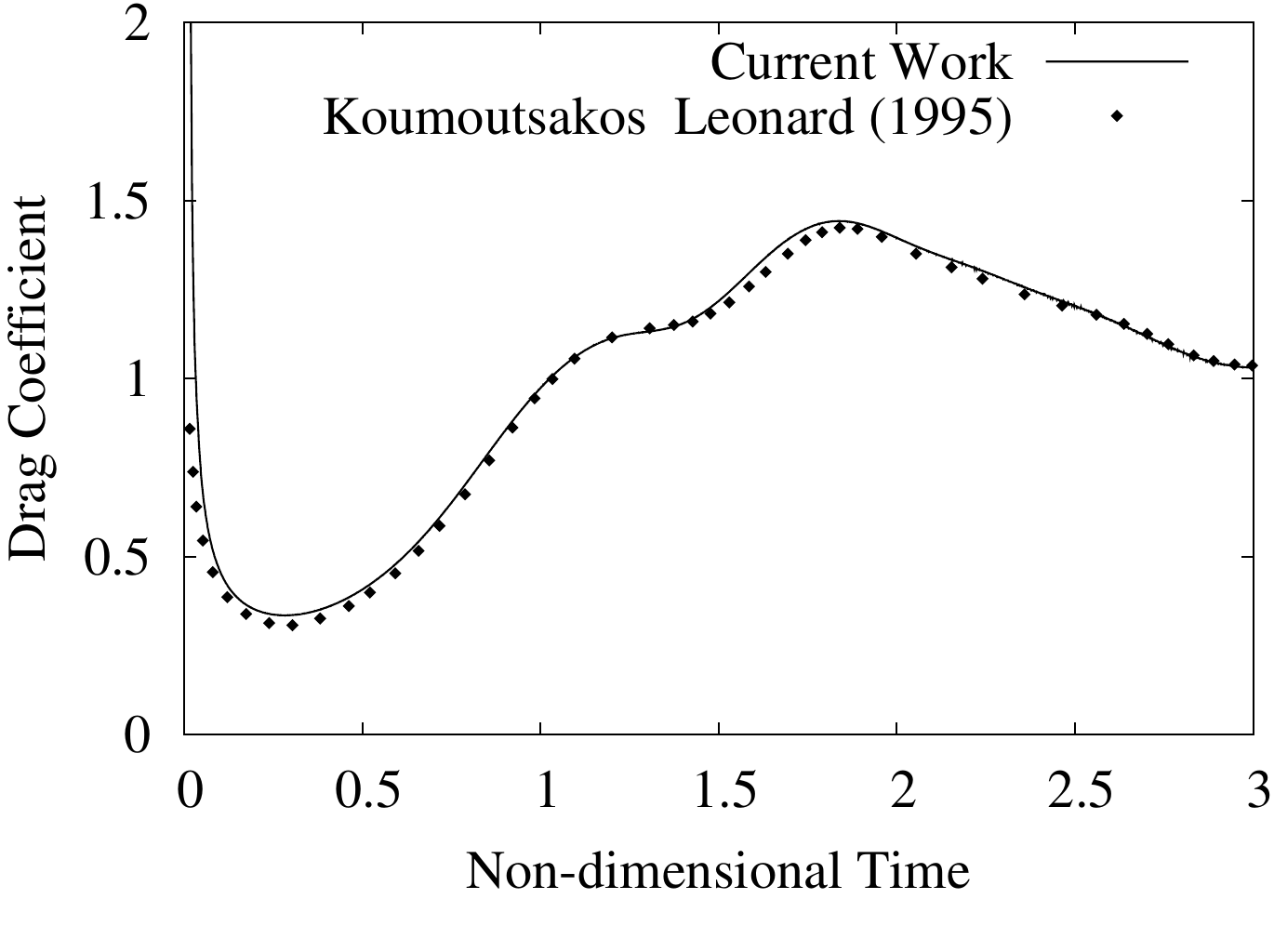}}
	\vspace{-3mm}\caption {\small (a) Vorticity field after non-dimensional time 3.0 and (b) time varying drag coefficient for external flow over a circular cylinder at Reynolds number 3000. The contour lines in (a) are drawn from -56 to 56 in steps of 4, excluding the zero contour.}
	\label{fig:Re3000}
\end{center}
\end{figure*}

The flow at $Re 40$ reaches a steady state and the drag coefficient was found to be 1.57, which matches the expected value \cite{Tritton1959}. The vorticity fields and the unsteady drag coefficients for the cases with $Re 550$ and $3000$ also agree well with past computations \cite{Koumoutsakos1995}. The results are presented in Figures \ref{fig:Re40}--\ref{fig:Re3000}.

\begin{table}[h]
\begin{center}
\small
    \begin{tabular}{ | c | c | c | c | c |}
	\hline
	Re & $n_x \times n_y$ & $\Delta{x}_{min}$ & Extent of uniform grid & $r_{stretching}$\\
	\hline
	$40$ & $330 \times 330$ & $0.02$ & $[-0.54,0.54]$ & $1.02$\\
	\hline
	$550$ & $450 \times 450$ & $0.01$ & $[-0.54,0.54]$ & $1.02$\\
	\hline
	$3000$ & $986 \times 986$ & $0.004$ & $[-0.52,0.52]$ & $1.01$\\
	\hline
    \end{tabular}
\end{center}
    \vspace{-4mm}\caption{\small Grid information for the impulsively started cylinder cases.}
    \label{table:gridinfo}
\end{table}

\subsection{External flow over a circular cylinder}

Longer runs were performed to simulate the von Karman street behind a circular cylinder. Table \ref{table:wakeresults} lists the results that were obtained for flows with different Reynolds numbers.

The cylinder is again of diameter 1 and in the center of a domain of size 30x30. The minimum cell width that was used for the runs was 0.02, with it being maintained for the whole distance behind the cylinder. The grid to the left, top and bottom of the cylinder is exponential with a stretching ratio of 1.02. The initial position of the cylinder is slightly offset in the y-direction and it is given a nudge to bring it to the center at the beginning of the run to trigger the instability in the flow and cause vortex shedding.

\begin{figure}[h]
\centering
	\includegraphics[width=0.48\textwidth]{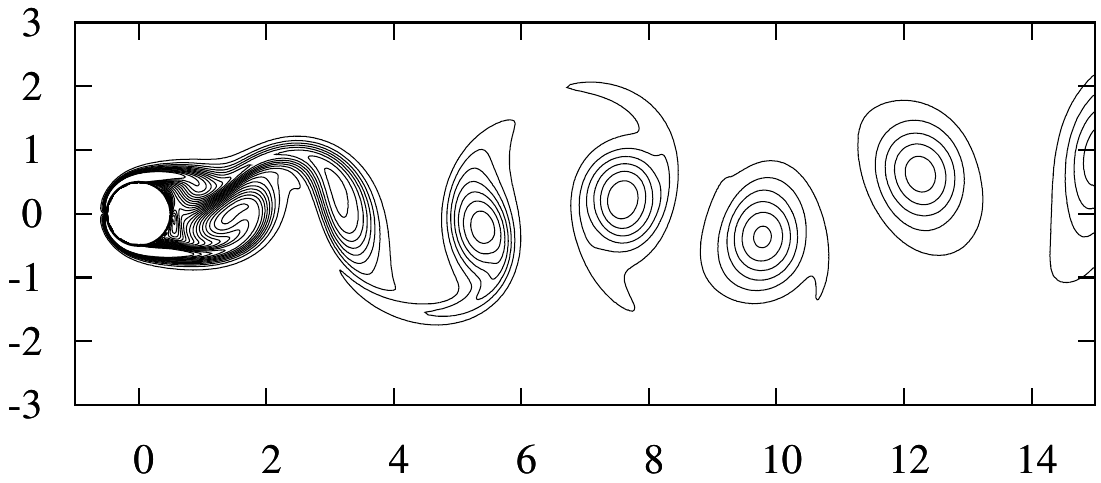}
	\vspace{-4mm}\caption{\small Flow over a circular cylinder at $Re$ $200$. Vorticity contours from -5 to 5 in increments of 0.4}
\end{figure}

\begin{table}
\centering
	\small
	\begin{tabular}{ | c | c | c | c | c |}
		\hline
		$Re$ & $C_l$ & $C_d$ & $St$ & $St_{experimental}$ \cite{Williamson1996}\\
		\hline
		$100$ & $\pm{0.339}$ & $1.37\pm{0.009}$ & $0.166$ & $0.164$\\
		\hline
		$150$ & $\pm{0.532}$ & $1.35\pm{0.026}$ & $0.185$ & $0.184$\\
		\hline
		$200$ & $\pm{0.688}$ & $1.36\pm{0.042}$ & $0.197$ & $0.197$\\
		\hline
	\end{tabular}
	\caption{\small Computed data for flow past a circular cylinder at different Reynolds numbers and comparison with experimental results}
	\label{table:wakeresults}
\end{table}

%

\section{Simulations of wake flows and moving boundaries}

The calculations presented above validate our code with both analytical results and experimental benchmarks, yet they are not useful to evaluate the performance of the \ibm\ when applied to moving boundary flows. With moving boundaries, the \ibm\ requires regenerating the matrix and the preconditioner at every time step (or every few time steps). For this reason, we include here some calculations that reproduce past results for flows with moving boundaries, for which we report the run times. All the runs were performed on an \NV\ Tesla C2050 {\gpu}. The first case is a heaving airfoil placed in a uniform flow (performed by Lewin and Haj-Hariri \cite{Lewin2003}) and the second is an airfoil that performs both pitching and heaving motions, simulating the flapping motion of an insect wing (performed by Wang et al \cite{Wang2004}).

\subsection{Heaving Airfoil}

\begin{figure*}
\begin{center}
	\subfloat[$t*=7.125$]{ \includegraphics[width=0.48\textwidth]{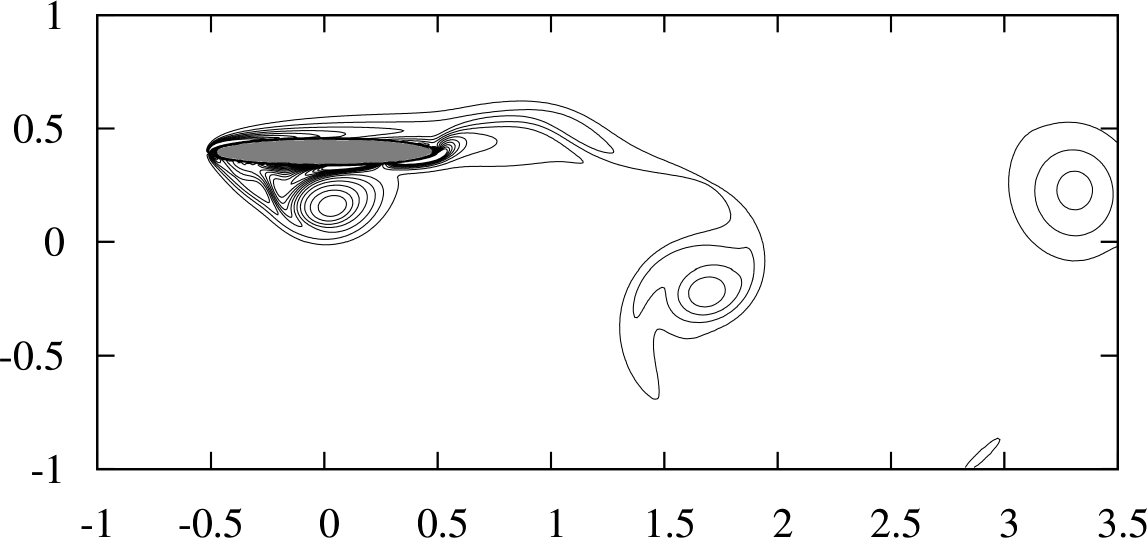} }
	\subfloat[$t*=7.3125$]{ \includegraphics[width=0.48\textwidth]{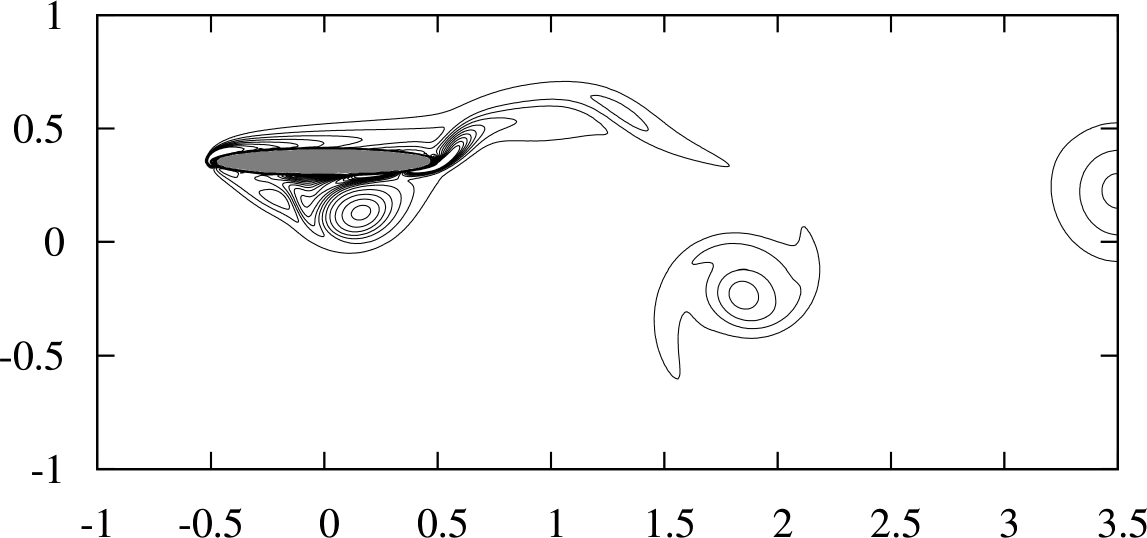} } \\
	\subfloat[$t*=7.5$]{ \includegraphics[width=0.48\textwidth]{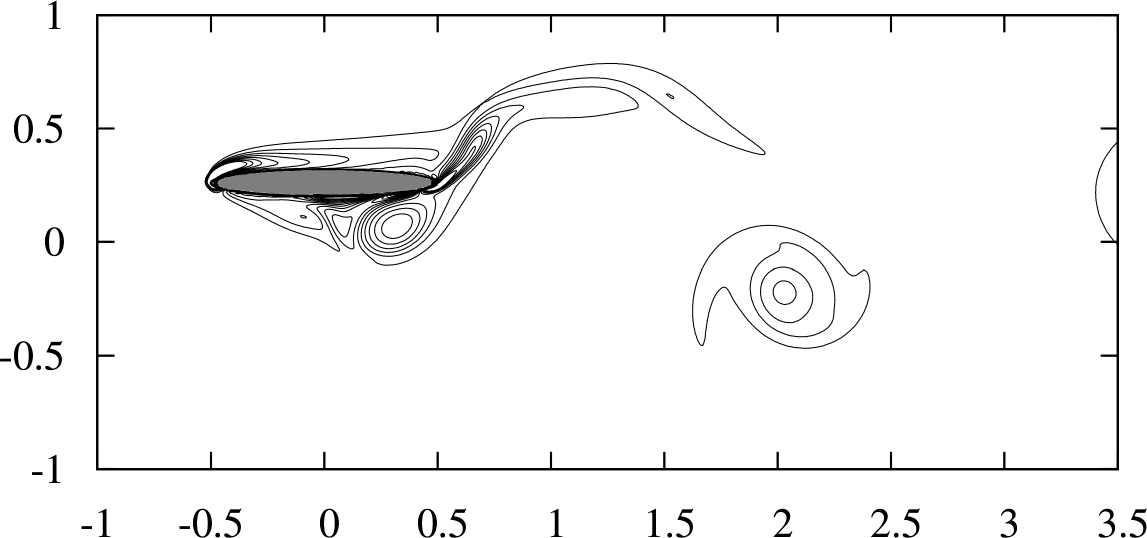} }
	\subfloat[$t*=7.6875$]{ \includegraphics[width=0.48\textwidth]{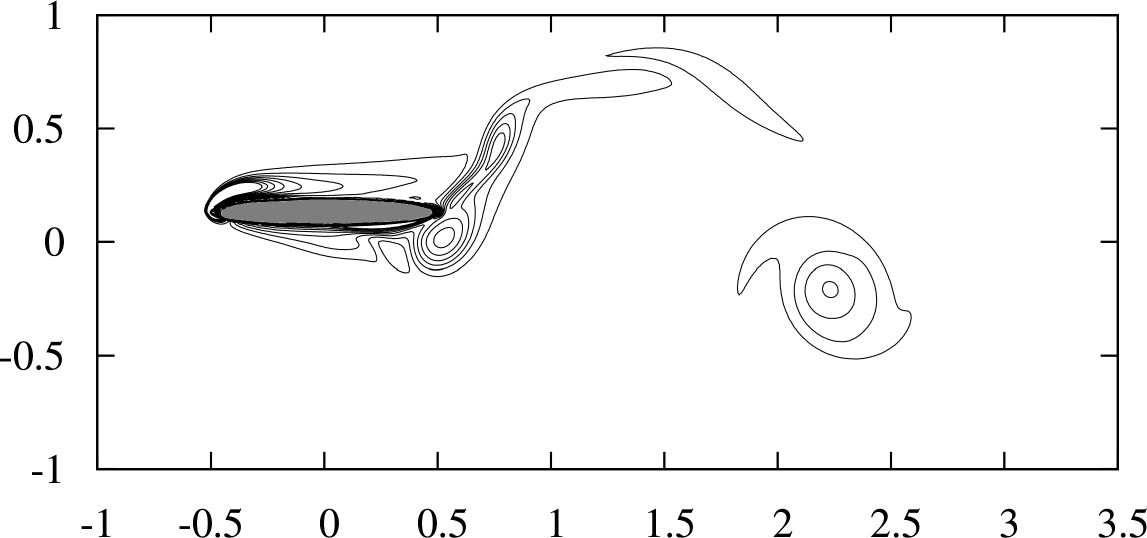} } \\
	\subfloat[$t*=8.0625$]{ \includegraphics[width=0.48\textwidth]{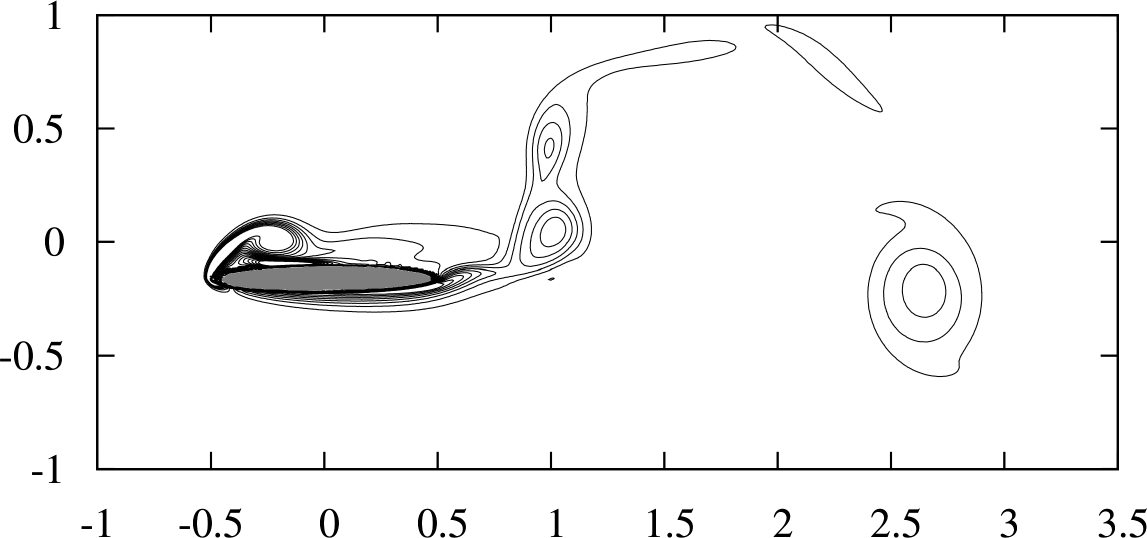} }
	\subfloat[$t*=8.4375$]{ \includegraphics[width=0.48\textwidth]{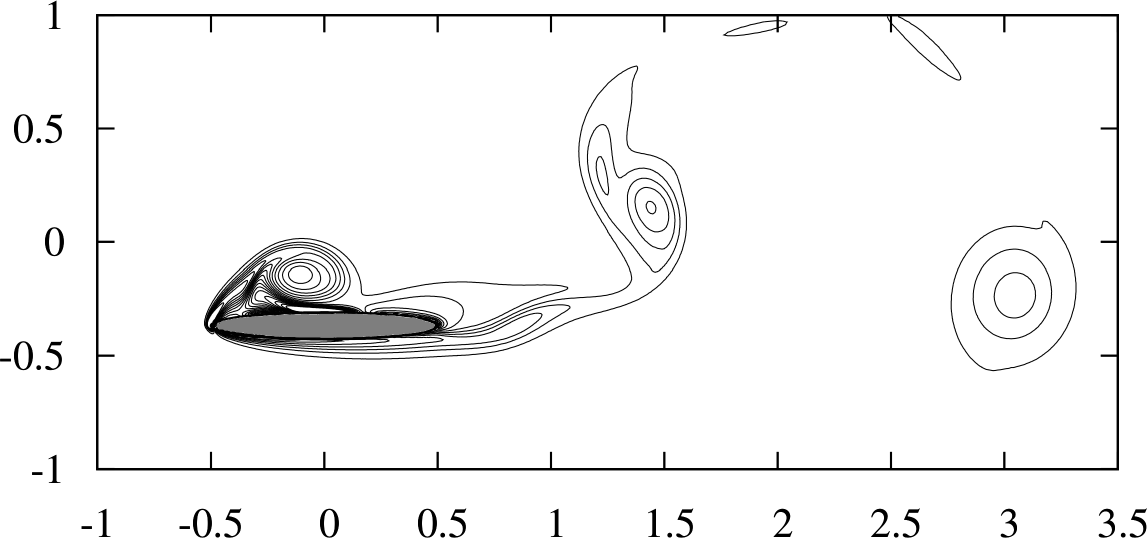} }
\caption{\small Vorticity field for the downstroke of a heaving airfoil in a flow with $Re=500$, $k=2.0$ and $kh=0.8$. Vorticity contours are drawn at levels $\pm{2}$, $\pm{6}$, $\pm{10}$, \emph{etc}. Compare with Figure 3 in \cite{Lewin2003}.}
\label{fig:heaving}
\end{center}
\end{figure*}

The simulation was carried out for an elliptic airfoil of thickness-to-chord ratio 0.12 and chord length $c=1$, heaving with a reduced frequency $k=2.0$ and non-dimensional maximum heaving velocity $kh=0.8$ at Reynolds number $Re=500$. The domain is of size $30 \times30$ and the near-body region in $[-0.52,0.52]\times[-0.52,0.52]$ has the minimum cell width of 0.005. The region $[0.52,0.78]$ immediately behind the airfoil has an exponential grid with stretching ratio 1.02, and a uniform grid of size 0.01 follows from that region to the edge of the domain. The grid in front of the airfoil is stretched with a ratio of 1.02 and the grid in the $y$-direction above and below the body is stretched with a ratio of 1.015. The total size of the mesh is $1339\times{686}$ and the time step used is 0.0005. The boundary conditions are the same as those used for the earlier cases of external flow over a cylinder. The obtained vorticity field (see Figure \ref{fig:heaving}) compares well with the results of Lewin and Haj-Hariri \cite{Lewin2003}. The linear solve to calculate the pressure and forces solves for over 900,000 unknowns and each time step of the simulation requires 5 seconds.

\subsection{Flapping airfoil}

We consider a flapping airfoil, the motion of which is described by:
\begin{align}
	x(t) = \frac{A_0}{2} \cos (2{\pi}{f}{t}) \nonumber \\
	\alpha (t) = \alpha_0 + \beta \sin ( 2{\pi}{f}{t} + \phi ), \nonumber
\end{align}
where $x(t)$ is the position of the center of the airfoil and $\alpha(t)$ is the angle made by the airfoil with the line of oscillation. The airfoil is elliptical with a thickness-to-chord ratio of 0.12 and rotates about its center. The Reynolds number is calculated using the maximum translational velocity of the airfoil and the chord length. We consider the case with symmetrical rotation $(\phi=0)$ at Reynolds number $75$ and with ${A_0}/{c}=2.8$, $\alpha=\pi/2$, $\beta=\pi/4$ and $f=0.25$ Hz.

The airfoil has chord length 1 and oscillates at the center of a domain, each side of which is 30 chord-lengths long. The grid is uniform in the region $[-2,2]\times[-0.52,0.52]$ with cell width 0.01 and beyond this region it is stretched with a ratio of 1.01 on all sides, resulting in a mesh size of $930\times{654}$ cells. The time step used is 0.001. The run time for 4000 time steps (one cycle) was 164 seconds, when the preconditioner was updated every two time steps.

\begin{figure*}
\begin{center}
$\begin{array}{cc}
\includegraphics[width=0.42\textwidth]{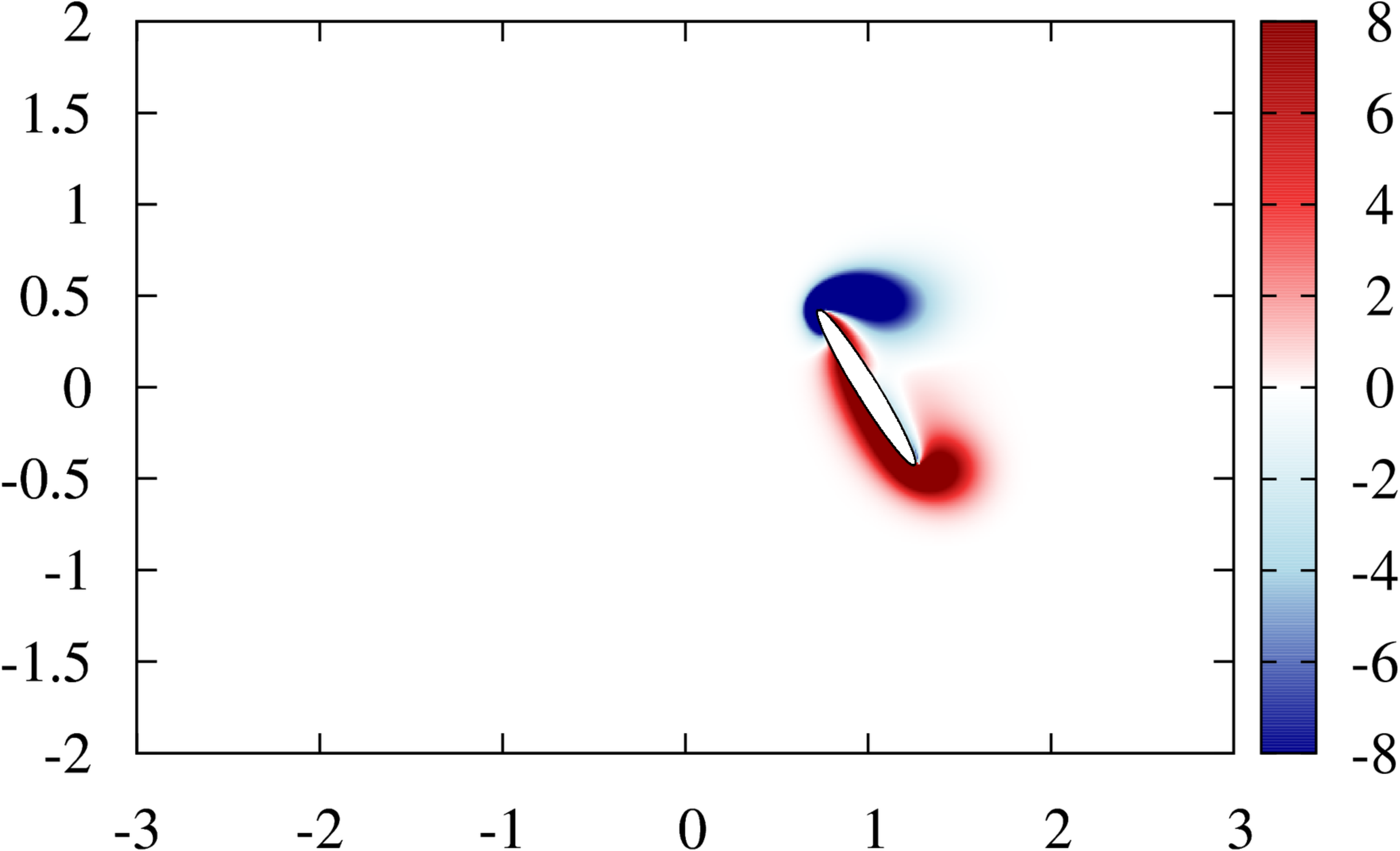} &
\includegraphics[width=0.42\textwidth]{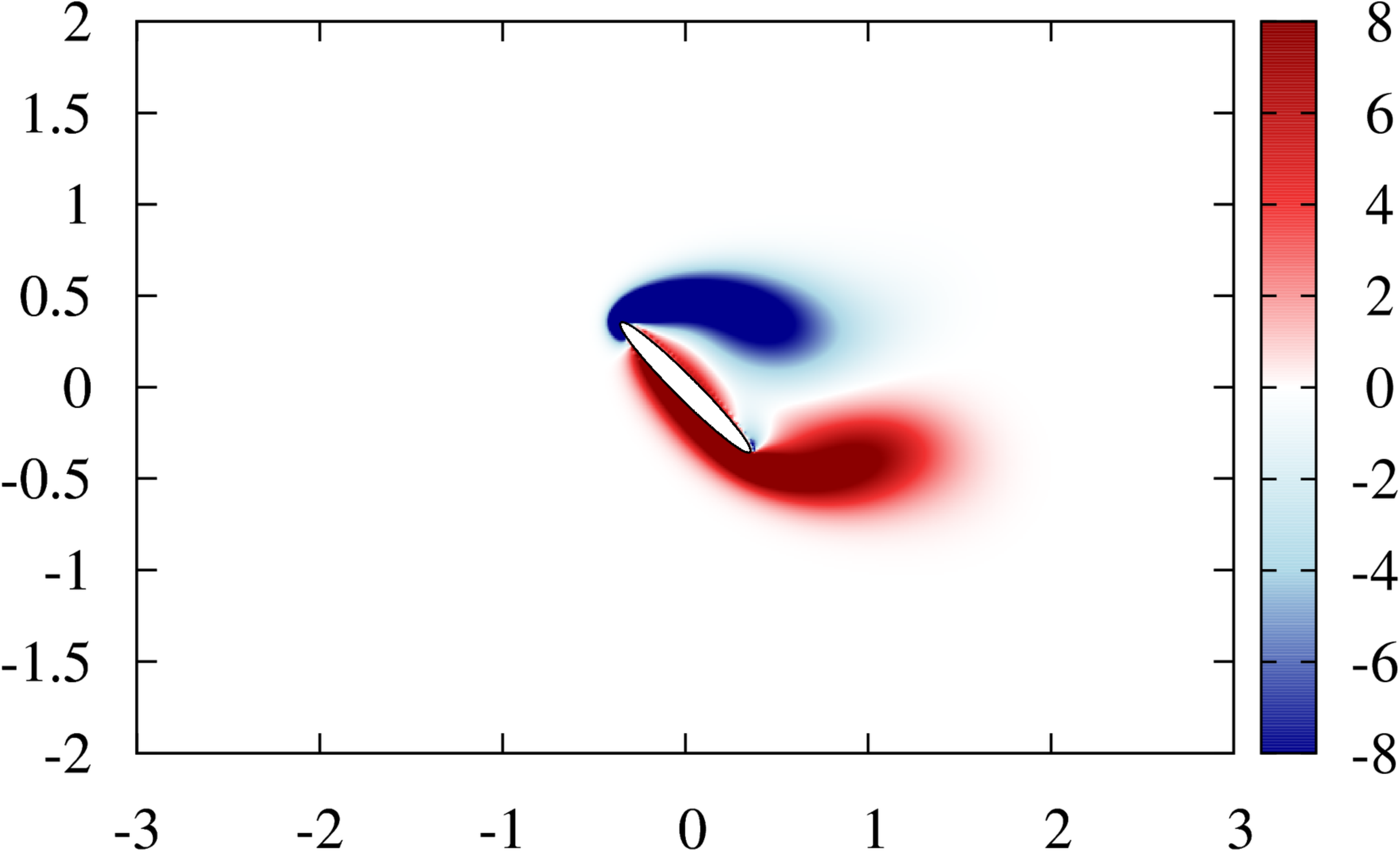} \\
\includegraphics[width=0.42\textwidth]{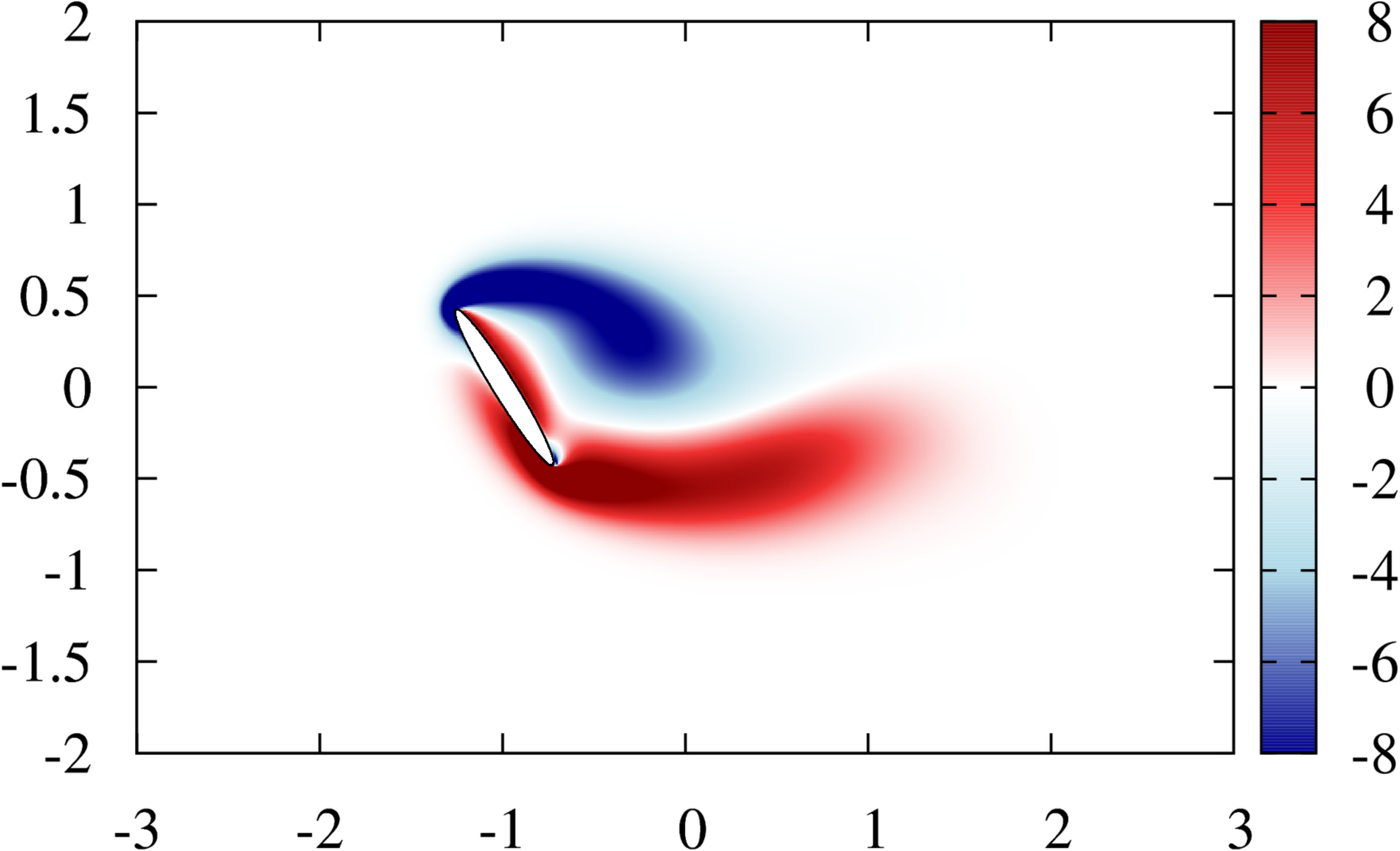} &
\includegraphics[width=0.42\textwidth]{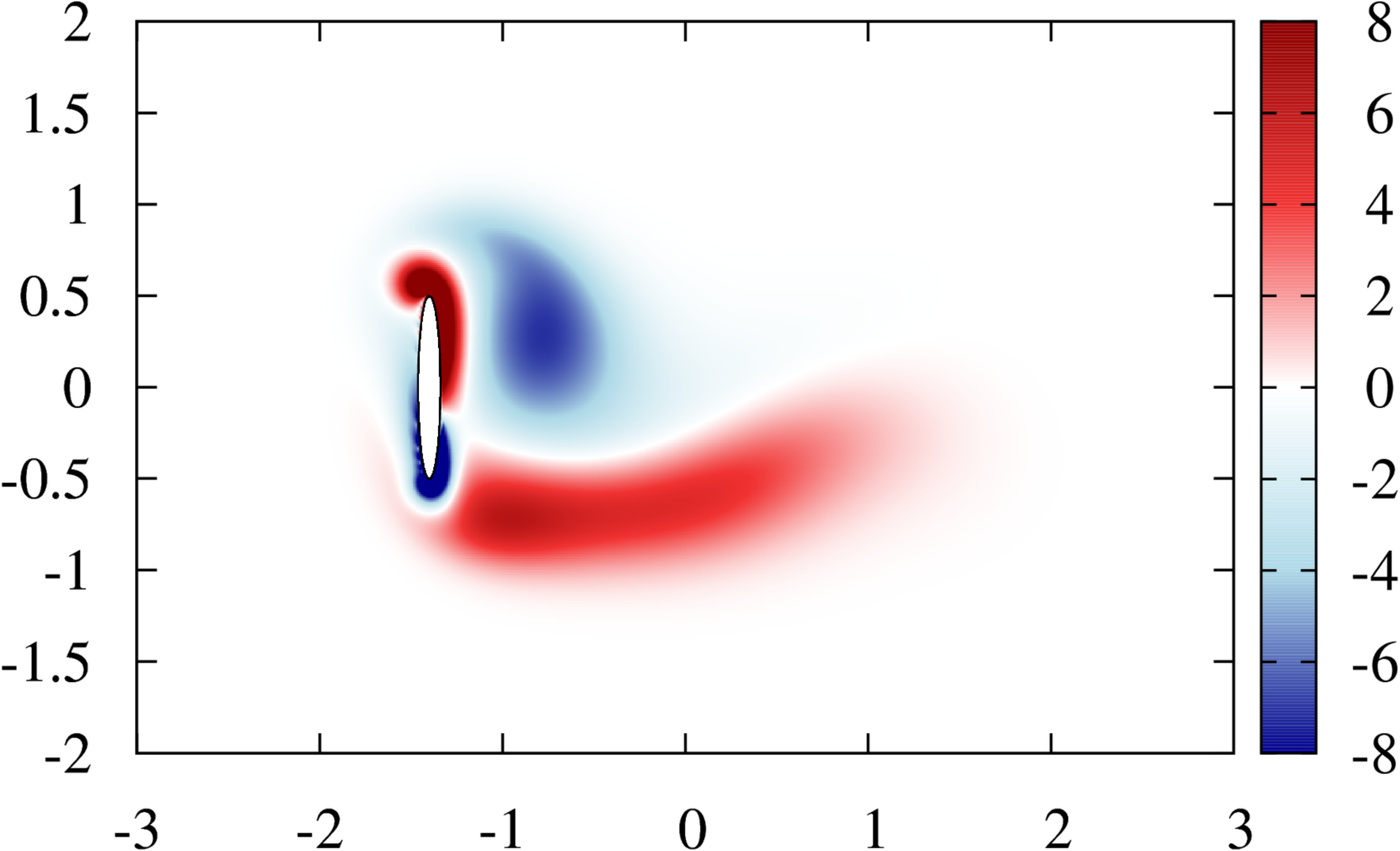} \\
\includegraphics[width=0.42\textwidth]{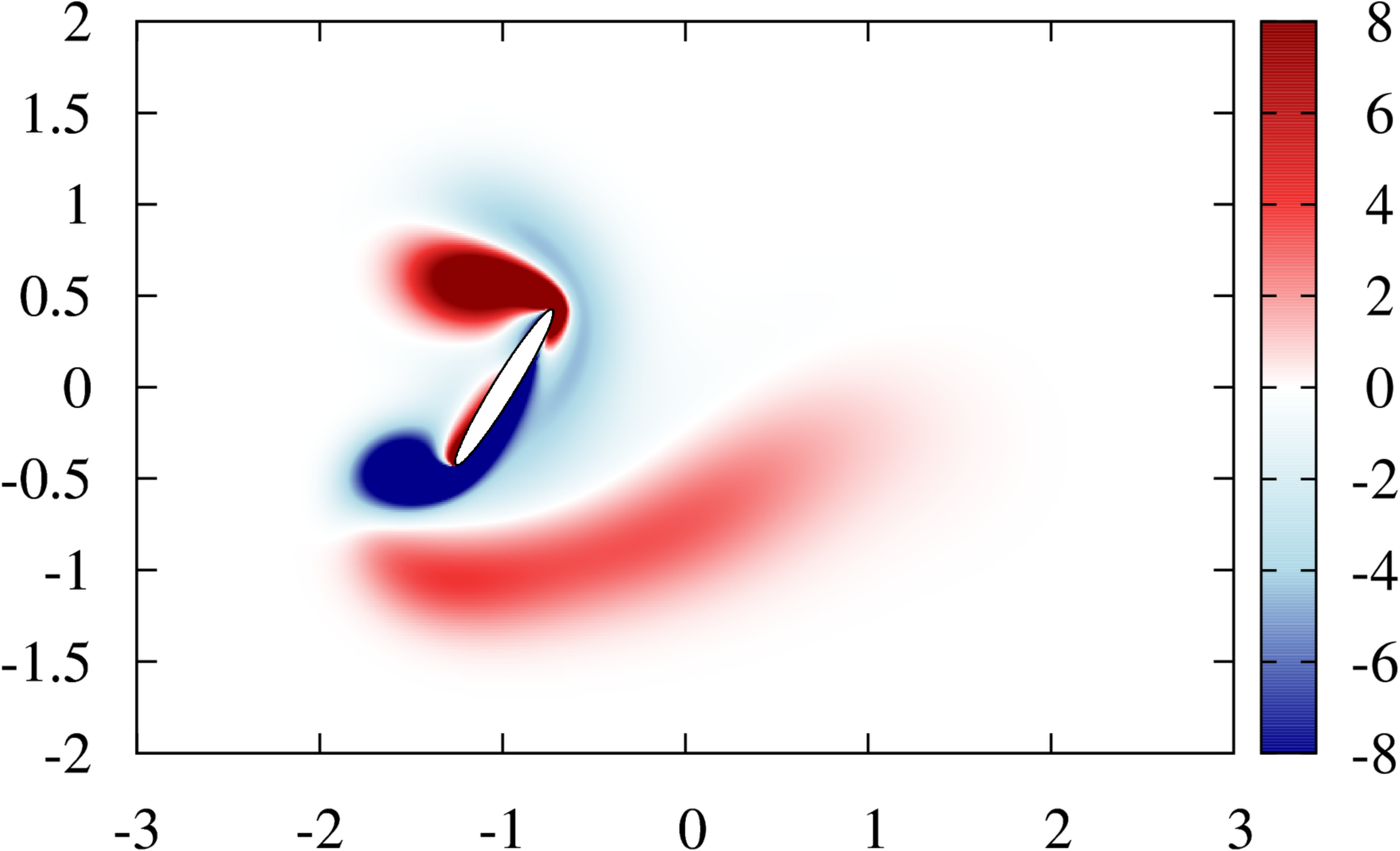} &
\includegraphics[width=0.42\textwidth]{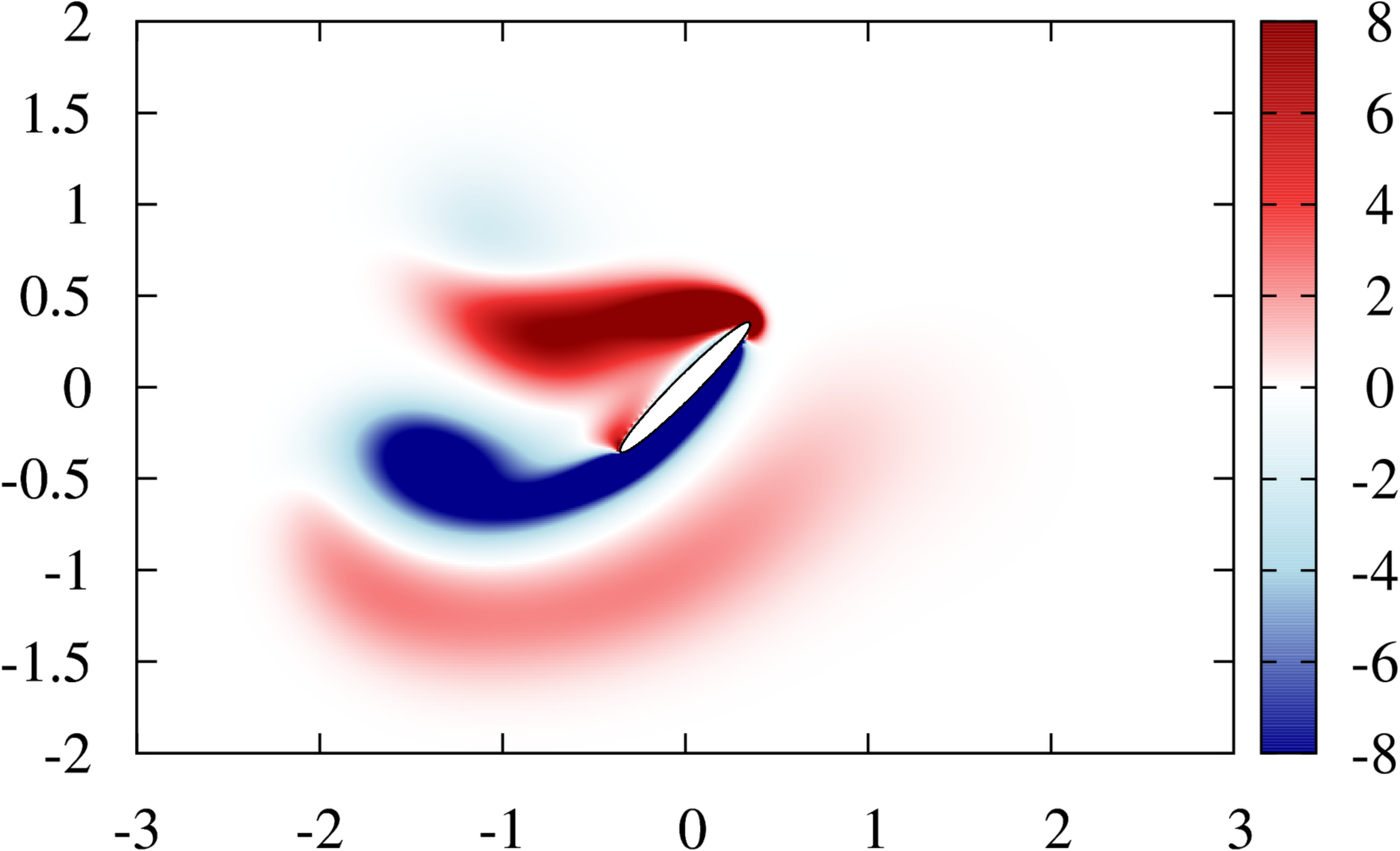} \\
\includegraphics[width=0.42\textwidth]{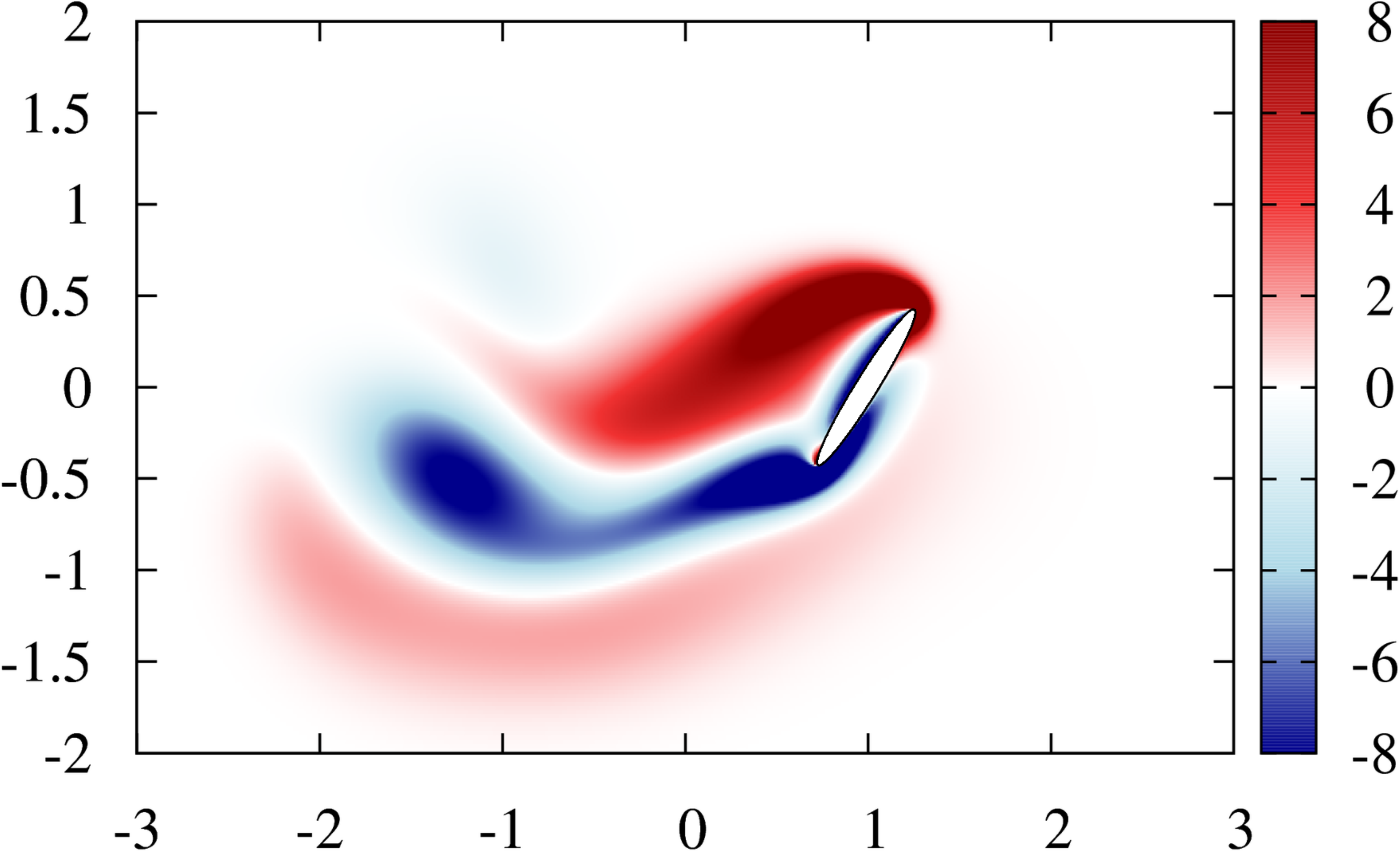} &
\includegraphics[width=0.42\textwidth]{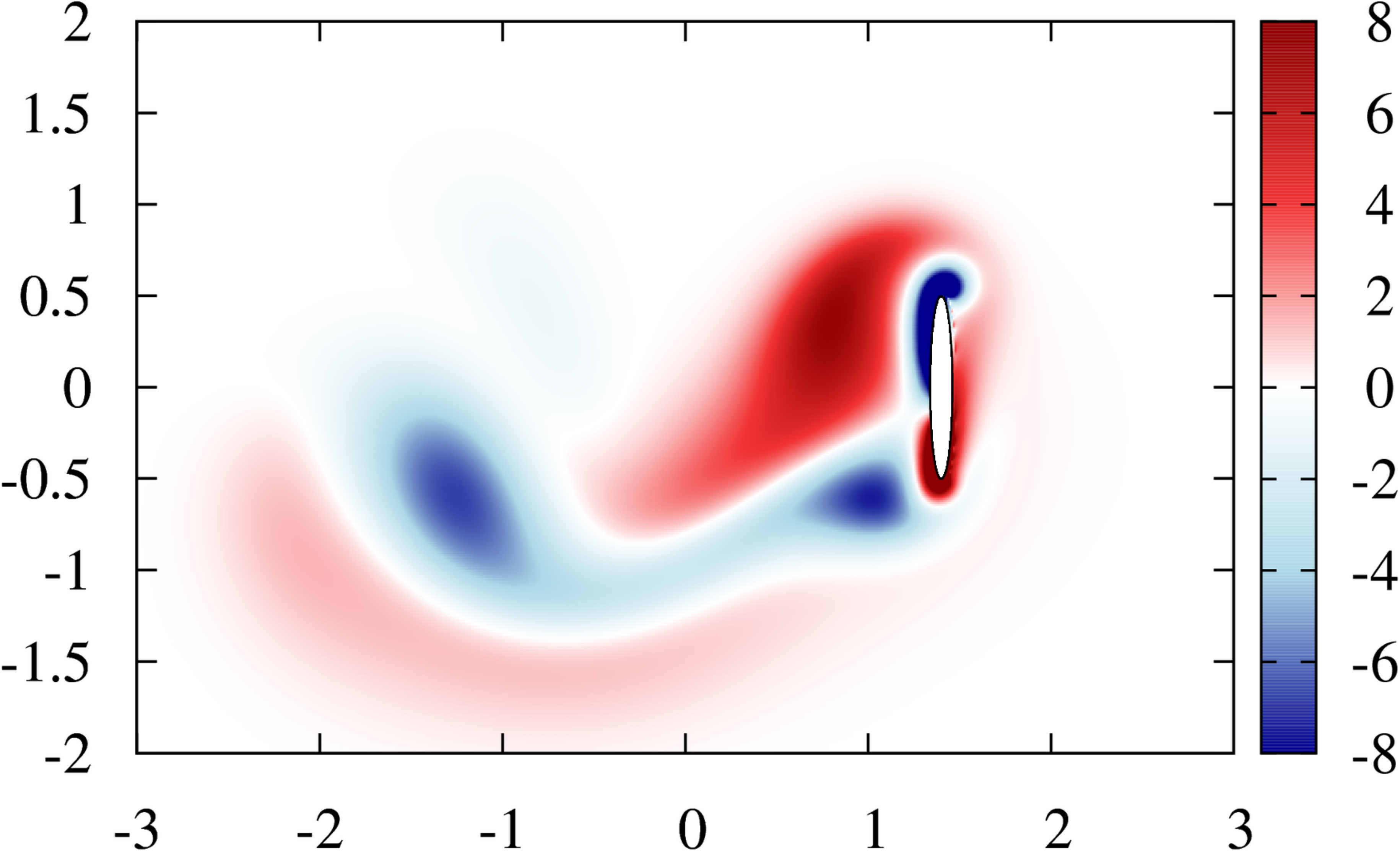}
\end{array}$
\caption{\small Vorticity field around a flapping airfoil with symmetric rotation ($\phi=0$, $A_0/c=2.8$) at Reynolds number 75. The frames represent equally spaced instances of time in the first cycle of flapping: $T=0.125$, $0.25$, $0.375$, \ldots (The time is non-dimensionalised by the time period of oscillation)}
\label{fig:flapping}
\end{center}
\end{figure*}

\begin{figure*}
\begin{center}
\includegraphics[angle=0,width=0.5\textwidth]{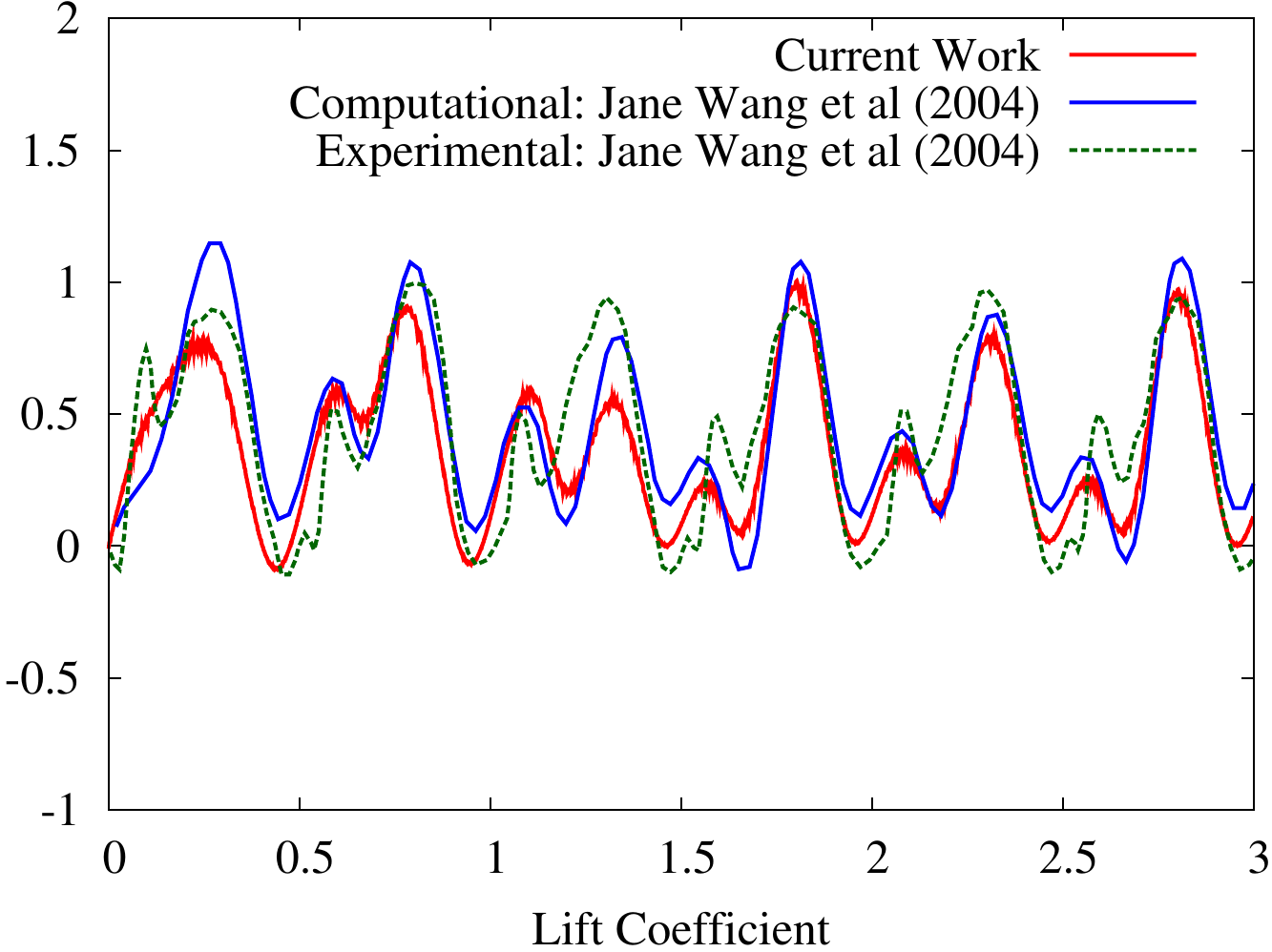}
\caption{\small Unsteady lift coefficient for the first three cycles of a flapping airfoil ($Re=75$, $\phi=0$, $A_0/c=2.8$). The time is non-dimensionalised using the time period of oscillation and the lift coefficient is calculated by normalising the lift force with respect to the maximum of the quasi-steady force experienced by the airfoil considered (see \cite{Wang2004} for more details)}
\label{fig:flapCl}
\end{center}
\end{figure*}

The vorticity field at different times during the first cycle is shown in Figure \ref{fig:flapping} and a comparison of the unsteady lift coefficient with the computational and experimental results presented by Wang et al \cite{Wang2004} is plotted in Figure \ref{fig:flapCl}. The experiments were conducted with a three-dimensional wing and both simulations were performed in two dimensions, hence we don't expect them to closely match. But we note that the computational results follow the expected trend and agree reasonably well with the experimental results.

\section{Conclusions and Future Work}

At this time, we have a validated \gpu\ code for the projection \ibm, and we have shown convergence with the expected rates. Using the free and open-source {\cusp} and \textsl{Thrust} libraries to provide sparse linear algebra functionality, we are able to quickly run large experiments to test a wide variety of problems including internal Couette flows, external flows past cylinders and complex moving bodies.


In this paper we have outlined a strategy for using the {\cusp} library effectively by reducing memory transfers between the host and device and where possible prioritizing the use of the most tuned routines. We save the limited memory available on the {\gpu} by using a modified routine to calculate a necessary triple matrix product without needing a costly intermediate matrix. Furthermore, we have investigated the suitability of a variety of linear solvers for our particular problem, demonstrating that for our problems a Conjugate Gradient method with a smoothed aggregation AMG preconditioner updated every 2 or 3 timesteps is the best available for reducing the total runtime of the algorithm.

We have demonstrated the capability of our code for both high Reynolds number flows and for flows with complex moving boundaries, comparing very well with existing works.

The next major step for this work, is to extend our algorithm into 3 dimensions---while this extension doesn't involve many differences to the equations, it will require a significant increase in memory usage. This will necessitate modifying the code to run in parallel on multiple {\gpu}s, a far more significant change. A further improvement would be to implement the {\ibm} with an adaptive grid -- this would reduce both the memory requirements and the amount of work needed per timestep.


\section{Acknowledgments}
We acknowledge support from NSF grant OCI-0946441, ONR award \#N00014-11-1-0356, and Boston University College of Engineering. LAB is also grateful of the support from NVIDIA via an Academic Partnership award.

\section*{References}
\small
\bibliographystyle{plain}
\bibliography{ibm,scicomp}

\end{document}